# The LHS 1678 System: Two Earth-sized Transiting Planets and an Astrometric Companion Orbiting an M Dwarf Near the Convective Boundary at 20 pc

Michele L. Silverstein[1,2,37], Joshua E. Schlieder[1], Thomas Barclay[1,3], Benjamin J. Hord[1,4], Wei-Chun Jao[2,5],
Eliot Halley Vrijmoet[2,5], Todd J. Henry[2], Ryan Cloutier[6,38], Veselin B. Kostov[1,7], Ethan Kruse[1,4],
Jennifer G. Winters[6], Jonathan M. Irwin[6], Stephen R. Kane[8], Keivan G. Stassun[9,10], Chelsea Huang[11,36],
Michelle Kunimoto[11], Evan Tey[11], Andrew Vanderburg[11], Nicola Astudillo-Defru[12], Xavier Bonfils[13], C. E. Brasseur[14],
David Charbonneau[6], David R. Ciardi[15], Karen A. Collins[6], Kevin I. Collins[16], Dennis M. Conti[17],
Ian J. M. Crossfield[18], Tansu Daylan[11,39], John P. Doty[19], Courtney D. Dressing[20], Emily A. Gilbert[1,3,21,22,23],
Keith Horne[14], Jon M. Jenkins[24], David W. Latham[6], Andrew W. Mann[25], Elisabeth Matthews[26],
Leonardo A. Paredes[2,5], Samuel N. Quinn[6], George R. Ricker[11], Richard P. Schwarz[27], Sara Seager[11,28,29],
Ramotholo Sefako[30], Avi Shporer[11], Jeffrey C. Smith[7,24], Christopher Stockdale[31], Thiam-Guan Tan[32,33],
Guillermo Torres[6], Joseph D. Twicken[7,24], Roland Vanderspek[11], Gavin Wang[34], and Joshua N. Winn[35]
[1] NASA Goddard Space Flight Center, Greenbelt, MD 20771, USA
[2] RECONS Institute, Chambersburg, PA 17201, USA
[3] University of Maryland, Baltimore County, 1000 Hilltop Circle, Baltimore, MD 21250, USA
[4] University of Maryland, College Park, MD 20742, USA
[5] Department of Physics and Astronomy, Georgia State University, 33 Gilmer Street SE Atlanta, GA 30303, USA
[6] Center for Astrophysics | Harvard & Smithsonian, 60 Garden Street, Cambridge, MA 02138, USA
[7] SETI Institute, 189 Bernardo Ave., Suite 200, Mountain View, CA 94043, USA
[8] Department of Earth and Planetary Sciences, University of California, Riverside, CA 92521, USA
[9] Vanderbilt University, Department of Physics & Astronomy, 6301 Stevenson Center Lane, Nashville, TN 37235, USA
[10] Fisk University, Department of Physics, 1000 18th Avenue N., Nashville, TN 37208, USA
[11] Department of Physics and Kavli Institute for Astrophysics and Space Research, Massachusetts Institute of Technology, Cambridge, MA 02139, USA
[12] Departamento de Matemática y Física Aplicadas, Universidad Católica de la Santísima Concepción, Alonso de Rivera 2850, Concepción, Chile
[13] Univ. Grenoble Alpes, CNRS, IPAG, F-38000 Grenoble, France
[14] SUPA Physics and Astronomy, University of St. Andrews, Fife, KY16 9SS, Scotland, UK
[15] Caltech/IPAC-NExScI, M/S 100-22, 1200 E. California Boulevard, Pasadena, CA 91125, USA
[16] George Mason University, 4400 University Drive, Fairfax, VA 22030, USA
[17] American Association of Variable Star Observers, 49 Bay State Road, Cambridge, MA 02138, USA
[18] Department of Physics and Astronomy, University of Kansas, Lawrence, KS 66045, USA
[19] Noqsi Aerospace Ltd., 15 Blanchard Avenue, Billerica, MA 01821, USA
[20] Department of Astronomy, University of California at Berkeley, Berkeley, CA 94720, USA
[21] Department of Astronomy and Astrophysics, University of Chicago, 5640 S. Ellis Avenue, Chicago, IL 60637, USA
[22] The Adler Planetarium, 1300 South Lakeshore Drive, Chicago, IL 60605, USA
[23] GSFC Sellers Exoplanet Environments Collaboration, Greenbelt, MD 20771, USA
[24] NASA Ames Research Center, Moffett Field, CA 94035, USA
[25] Department of Physics and Astronomy, University of North Carolina at Chapel Hill, Chapel Hill, NC 27599, USA
[26] Observatoire de lUniversité de Genève, Chemin Pegasi 51,1290 Versoix, Switzerland
[27] Patashnick Voorheesville Observatory, Voorheesville, NY 12186, USA
[28] Department of Earth, Atmospheric, and Planetary Sciences, Massachusetts Institute of Technology, Cambridge, MA 02139, USA
[29] Department of Aeronautics and Astronautics, MIT, 77 Massachusetts Avenue, Cambridge, MA 02139, USA
[30] South African Astronomical Observatory, P.O. Box 9, Observatory, Cape Town 7935, South Africa
[31] Hazelwood Observatory, VIC, Australia
[32] Perth Exoplanet Survey Telescope, Perth, Western Australia, Australia
[33] Curtin Institute of Radio Astronomy, Curtin University, Bentley, Western Australia 6102, Australia
[34] Tsinghua International School, Beijing 100084, People's Republic of China
[35] Department of Astrophysical Sciences, Princeton University, 4 Ivy Lane, Princeton, NJ 08544, USA
[36] Centre for Astrophysics, University of Southern Queensland, Toowoomba, QLD, 4350, Australia
*Received 2021 June 24; revised 2021 October 20; accepted 2021 October 22; published 2022 March 7*

## Abstract

We present the Transiting Exoplanet Survey Satellite (TESS) discovery of the LHS 1678 (TOI-696) exoplanet system, comprised of two approximately Earth-sized transiting planets and a likely astrometric brown dwarf orbiting a bright ($V_J = 12.5$, $K_s = 8.3$) M2 dwarf at 19.9 pc. The two TESS-detected planets are of radius $0.70 \pm 0.04\,R_\oplus$ and $0.98 \pm 0.06\,R_\oplus$ in 0.86 day and 3.69 day orbits, respectively. Both planets are validated and characterized via ground-based follow-up observations. High Accuracy Radial Velocity Planet Searcher RV monitoring yields 97.7 percentile mass upper limits of 0.35 $M_\oplus$ and 1.4 $M_\oplus$ for planets b and c, respectively. The astrometric companion detected by the Cerro Tololo Inter-American Observatory/Small and Moderate Aperture Telescope System 0.9 m has an orbital period on the order of decades and is undetected by other means. Additional ground-based observations constrain the companion to being a high-mass brown dwarf or smaller. Each planet is of unique interest; the inner planet has an ultra-short period, and the outer planet is in the Venus zone. Both are promising targets for atmospheric characterization with the James Webb Space Telescope and mass measurements via extreme-precision radial velocity. A third planet candidate of radius $0.9 \pm 0.1\ R_\oplus$ in a 4.97 day orbit is also identified in multicycle TESS data for validation in future work. The host star is associated with an observed gap in the lower main sequence of the Hertzsprung–Russell diagram. This gap is tied to the transition from partially to fully convective interiors in M dwarfs, and the effect of the associated stellar astrophysics on exoplanet evolution is currently unknown. The culmination of these system properties makes LHS 1678 a unique, compelling playground for comparative exoplanet science and understanding the formation and evolution of small, short-period exoplanets orbiting low-mass stars.

---

[37] NASA Postdoctoral Program Fellow.
[38] Banting Fellow.
[39] Kavli Fellow.

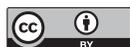

*Unified Astronomy Thesaurus concepts:* Exoplanet systems (484); Transit photometry (1709); Low mass stars (2050); M dwarf stars (982); Astrometric binary stars (79)

## 1. Introduction

Small planets ($<4\ R_\oplus$) are frequently found around low-mass stars, which are in turn the most common stars in the Galaxy (Henry et al. 2006). The preponderance of small planets on short-period orbits around M dwarfs is revealed in statistical analyses of both ground-based radial velocity survey discoveries (Bonfils et al. 2013) and transiting planets from the Kepler survey (Dressing & Charbonneau 2013, 2015; Hardegree-Ullman et al. 2019). High-priority individual systems discovered using dedicated ground-based transit surveys (e.g., from MEarth, Irwin et al. 2009, and TRAPPIST, Jehin et al. 2011) added to this haul of planetary systems. These systems, with small host stars and relatively deep planet transits, present some of the best opportunities for detailed characterization of small-planet bulk compositions and atmospheres.

This M dwarf advantage led to these stars being a focus of exoplanet searches with the Transiting Exoplanet Survey Satellite (TESS; Ricker et al. 2015), although it is not the sole purpose of the mission. TESS is performing a nearly all-sky survey to identify exoplanet systems around the nearest, brightest stars. These are most compelling for follow-up characterization studies to measure masses and investigate atmospheres with, e.g., the Hubble Space Telescope (HST) and the James Webb Space Telescope (JWST). M dwarfs are brightest in the red-optical to near-infrared wavelength range, though still fainter than more massive main sequence stars. With this in mind, TESS uses a broadband filter that ranges from 600 to 1000 nm, the red end of silicon CCD sensitivity. With ongoing discoveries from TESS and its improved sensitivity to these ubiquitous stars, our ability to focus on characterizing these exoplanet systems is improved compared to the Kepler mission.

Scattered among these bright nearby systems discovered by TESS are those uniquely poised for interesting follow-up observations. These include some of the closest and brightest M dwarf systems: LTT 1445 A (Winters et al. 2019), GJ 357 (Luque et al. 2019), and AU Microscopii (Plavchan et al. 2020). TESS has also revealed small planets orbiting bright M dwarfs with extreme orbital architectures, like the ultra-short-period LHS 3844 (Vanderspek et al. 2019) in an 11 hr orbit. Kreidberg et al. (2019) demonstrated the feasibility of characterizing such a planet's atmosphere, or lack thereof, by measuring the thermal emission coming from LHS 3844 b using Spitzer. M dwarf multiplanet systems amenable to detailed characterization have also been uncovered by TESS. Both the L 98-59 (Kostov et al. 2019a) and LP 791-18 (Crossfield et al. 2019) systems are comprised of multiple planets of different sizes orbiting bright nearby M dwarfs. These systems include small, relatively cool planets that are key for detecting atmospheres in transit. Planets in both systems are priority targets for early science observations with the JWST.

Continuing this stream of M dwarf planet discoveries from TESS is the LHS 1678 system. Here we present a holistic characterization that includes an in-depth analysis of the host star's properties, the identification of a wide-orbit, low-mass companion, and the characterization of two Earth-sized planets and a third small planet candidate in a compact system.

## 2. TESS Observations, Candidate Identification, and Vetting

LHS 1678 (TIC 77156829, TOI-696, L 375-2, LTT 2022, NLTT 13515) was observed by TESS in Sector 4 from UT 2018 October 19 to November 14 (25.95 days), in CCD 1 of Camera 3, and in Sector 5 from UT 2018 November 15 to December 11 (26.31 days), in CCD 2 of Camera 3. LHS 1678 data were collected at 2 minutes cadence because the star was prioritized for high-cadence measurements as part of the Cycle 1 Guest Investigator Program G011180,[40] the Cool Dwarf target catalog (Muirhead et al. 2018),[41] and the TESS Candidate Target List (CTL; Stassun et al. 2018b). LHS 1678 was also observed in Sectors 31 and 32 at 2 minutes cadence, after the majority of the analyses here were performed. We briefly discuss preliminary findings from the full four-sector TESS data set in Section 7.1. Aside from the aforementioned section, we focus our analyses on TESS Sectors 4 and 5 data only.

TESS 2 minute cadence data were processed into target pixel files (TPFs) and subsequent light curves by the NASA Ames Science Processing Operations Center pipeline (SPOC; Jenkins et al. 2016). The light curves revealed a star with little variability above the TESS noise level (indicating a lack of surface activity such as star spots) over nearly two months of continuous observations and exhibited no evidence for flares. The light curves were searched for periodic, transit-like signals using the Transiting Planet Search module (TPS; Jenkins et al. 2010) and revealed two candidate planet signals that passed a series of data validation tests (Twicken et al. 2018; Li et al. 2019). The TESS pipeline identified two candidates, TOI-696.01 and .02, with periods of 0.86 and 14.76 days, respectively, and transit depths that corresponded to planets with radii approximately 1 $R_\oplus$ for an early M dwarf host star (see Section 3). Based on this, we began independent vetting and validation studies of the candidates to determine if the system should be prioritized for further follow-up.

---

[40] Differential Planet Occurrence Rates for Cool Dwarfs – PI: C. Dressing. Details of approved TESS Guest Investigator Programs are available from https://heasarc.gsfc.nasa.gov/docs/tess/approved-programs.html.
[41] http://vizier.u-strasbg.fr/viz-bin/VizieR?-source=J/AJ/155/180

Following our team's established procedures (Kostov et al. 2019a; Gilbert et al. 2020) we used the DAVE vetting tool (Kostov et al. 2019b) to identify potential sources of false positives and verify the results of the SPOC pipeline validation tests. We verified that the transits associated with TOI-696.01 were on-target, and there were no differences in odd- and even-numbered transits. The phase-folded sets of odd- and even-numbered transits showed similar depth and shape, and there was no evidence for a secondary eclipse indicative of an eclipsing binary. Our DAVE vetting of TOI-696.02 was less straightforward. There were only two transits per sector at the 14.76 day SPOC-identified period, so centroid estimates and odd/even checks were not meaningful. However, the analysis did reveal additional transit-like signals at 1/4 the period identified by the TPS pipeline. We also used preliminary host star parameters as input to the publicly available statistical validation tool vespa (Morton 2015) to estimate the numerical probability that the identified candidates were false positives (see also Section 6). The vespa analysis indicated a low false-positive probability (FPP) for TOI-696.01 (∼1%) but a high probability (∼75%) that TOI-696.02 at the 14.76 day period was a false positive, resulting from the residual transit-like signals also identified in the DAVE vetting.

Following this lead, we examined the SPOC data validation reports for TOI-696.02 and found similar evidence for excess signal at 1/4 the originally identified period in the phased light curve. We then performed two independent planet candidate searches in an attempt to recover the TPS candidates at the same periods. Using both a modified version of the Quasi-periodic Automated Transit Search (QATS; Carter & Agol 2013; Kruse et al. 2019) and Transit Least Squares (TLS; Hippke & Heller 2019a, 2019b) search pipelines, we identified transiting planet candidates with periods of 0.86 days and 3.69 days, recovering TOI-696.01 at the same period and identifying TOI-696.02 at one-quarter the original period, consistent with the excess signal identified in our vetting and validation efforts.

With the SPOC team, we investigated the details of the TPS candidate search to reconcile the period difference for TOI-696.02. The discrepancy was traced to statistical tests and threshold settings in TPS (Seader et al. 2013) that were failed (but only marginally) by TOI-696.02 at the shorter period. While the multiple event statistic (MES, a detection statistic tuned to measure the detection strength of periodic transit-like signals) for the candidate was higher at a period of 3.69 days, there were more individual transits and their contributions to the MES were not sufficiently consistent to pass additional statistical tests in TPS. This resulted in a period of 14.76 days being favored, where fewer individual transits contributed to the MES, but in a more consistent way. With fine tuning, TPS does recover the candidate with higher MES at 3.69 days, consistent with our independent QATS and TLS searches. We verified that the shorter 3.69 day period for TOI-696.02 produced reliable and consistent results in reanalyses using both DAVE and vespa. For the subsequent analyses presented in this paper we adopt periods of 0.86 and 3.69 days, respectively, for the planet candidates TOI-696.01 and .02.

## 3. The Host Star

### 3.1. Observed Parameters

We report measured astrometric, kinematic, photometric, and spectroscopic properties of LHS 1678 in Table 1, both taken from the literature and observed in this work.

Literature astrometry data come from Gaia Data Release 2 (DR2; Gaia Collaboration et al. 2016, 2018; Lindegren et al. 2018), the most current release at the time of our analyses. The most recent Gaia Early Data Release 3 (EDR3; Lindegren et al. 2021) parallax is only 0.1% different; an update to EDR3 is not expected to have significant scientific return. Photometric data come from Gaia DR2, Winters et al. (2015), the TESS Input Catalog Version 8 (TIC v8; Stassun et al. 2019), the Two Micron All-Sky Survey (2MASS; Cutri et al. 2003; Skrutskie et al. 2006), and the Wide-field Infrared Survey Explorer (WISE) AllWISE data release (Wright et al. 2010; Cutri et al. 2014). The *BP* and *RP* bandpasses are slightly different in Gaia EDR3 (Riello et al. 2021), showing up as a 0.02 magnitude difference for LHS 1678. We opt to use the DR2 photometry for comparison to previous works that use the DR2 magnitudes (e.g., in Section 7.5). Radial velocities (RVs) come from observations using the CHIRON spectrograph on the Small and Moderate Aperture Telescope System (SMARTS) 1.5 m at the Cerro Tololo Inter-American Observatory (CTIO, CTIO/SMARTS 1.5m) and High Accuracy Radial Velocity Planet Searcher (HARPS) spectrograph on the European Southern Observatory (ESO) 3.6 m. The details of the RV observations and analysis are described in the context of system follow-up in Section 4.3. Kinematic information is derived by combining astrometric information and RVs. Our methodology is described in the context of possible thick disk membership in Section 3.3. Spectral types come from Reid et al. (2007) and the Pecaut & Mamajek (2013) color-temperature table.[42] We complement these results with our own, derived following the procedure of Henry et al. (2002) using a spectrum taken on 2005 January 30, using the CTIO/SMARTS 1.5 m Ritchey–Chrétien Spectrograph (RC Spec; Figure 1). These methods yield results within 0.5 subtypes of each other at M2.0 to M2.5.

Photometric metallicity relations from Bonfils et al. (2005; their Equation (1); $\mathcal{M}_K$, $V$, $K$), Mann et al. 2013 (their Equation (29); $J$, $K$ as per $V - K < 5.5$), and Kesseli et al. (2019; their Equation (6); $J$, $K$) yield estimated [Fe/H] values of −0.54, −0.36, and −0.67, respectively, for LHS 1678. As we will discuss in the context of stellar age in Section 3.3, these are in line with the star's position on the Hertzsprung–Russell (HR) diagram, spectral energy distribution (SED) fitting results, etc., all of which imply its metallicity is lower than most main sequence stars with similar effective temperature. Without an appropriate spectrum from a medium-resolution optical/near-infrared spectrograph with an established metallicity pipeline for M dwarfs, which includes TripleSpec on the Palomar Hale 200″ telescope (Rojas-Ayala et al. 2012; Dressing et al. 2017), SNIFS on the University of Hawai'i 88″ telescope (i.e., Mann et al. 2013), and SpeX on the NASA Infrared Telescope Facility 3 m (i.e., Mann et al. 2013), we do not adopt a

---

[42] http://www.pas.rochester.edu/~emamajek/EEM_dwarf_UBVIJHK_colors_Teff.txt

Table 1
Host Star Observed Properties and Literature Values

| Property | Value | Error | Ref. |
|---|---|---|---|
| Identifiers | LHS 1678, TIC 77156829, TOI-696, L 375-2, LTT 2022 NLTT 13515, Gaia DR2 4864160624337973248 | | |
| Astrometry and Kinematics | | | |
| R.A. J2015.5 (deg) | 068.17898662887 | 0.0195 | Gaia DR2 |
| decl. J2015.5 (deg) | −39.79087441181 | 0.0253 | Gaia DR2 |
| Parallax (mas) | 50.2773 | 0.0236 | Gaia DR2 |
| R.A. Proper motion (mas yr$^{-1}$) | 239.410 | 0.041 | Gaia DR2 |
| decl. Proper motion (mas yr$^{-1}$) | −967.772 | 0.051 | Gaia DR2 |
| Radial velocity (km s$^{-1}$) | 11.4667 | 0.0259 | This work (HARPS+CHIRON) |
| $U_{LSR}$ (km s$^{-1}$) | 81.0 | 0.3 | This work |
| $V_{LSR}$ (km s$^{-1}$) | −47.2 | 0.4 | This work |
| $W_{LSR}$ (km s$^{-1}$) | 12.8 | 0.3 | This work |
| Total galactic motion (km s$^{-1}$) | 94.6 | 0.4 | This work |
| Photometry | | | |
| NUV | 21.546 | 0.460 | GALEX |
| BP | 12.7467 | 0.0020 | Gaia DR2 |
| RP | 10.4688 | 0.0013 | Gaia DR2 |
| G | 11.5403 | 0.0005 | Gaia DR2 |
| U | 15.230 | ⋯ | Mermilliod (2006) |
| $V_J$ | 12.48 | 0.03 | Winters et al. (2015) |
| $R_{KC}$ | 11.46 | 0.03 | Winters et al. (2015) |
| $I_{KC}$ | 10.26 | 0.03 | Winters et al. (2015) |
| T | 10.4276 | 0.0073 | TIC v8 |
| J | 9.020 | 0.032 | 2MASS |
| H | 8.501 | 0.047 | 2MASS |
| $K_s$ | 8.265 | 0.029 | 2MASS |
| W1 | 8.099 | 0.024 | AllWISE |
| W2 | 7.947 | 0.020 | AllWISE |
| W3 | 7.870 | 0.018 | AllWISE |
| W4 | 7.783 | 0.134 | AllWISE |
| Spectral Characteristics | | | |
| Spectral type | M2.0V | 0.5 | This work: RC Spec |
| | M2.5 | 1.0 | Pecaut & Mamajek (2013) relations[a] |
| | M2.0 | ⋯ | Reid et al. (2007) |
| Metallicity ([Fe/H]) | subsolar | ⋯ | This work: HR diagram, SED fits, etc. (Sections 3.2, 3.3) |
| | −0.54 | 0.20 | Bonfils et al. (2005) relation |
| | −0.36 | 0.06 | Mann et al. (2013) relation |
| | −0.67 | 0.41 | Kesseli et al. (2019) relation |
| log $R'_{HK}$ | −6.087 | 0.548 | This work |
| | −6.08 | ⋯ | Rains et al. (2021) |

**Notes.** Gaia DR2: Gaia Collaboration et al. (2016, 2018); Lindegren et al. (2018); Galaxy Evolution Explorer (GALEX): Martin et al. (2005); Bianchi et al. (2011); TIC v8: Stassun et al. (2019); 2MASS: Cutri et al. (2003); Skrutskie et al. (2006); WISE AllWISE Release: Wright et al. (2010); Cutri et al. (2014).
[a] http://www.pas.rochester.edu/~emamajek/EEM_dwarf_UBVIJHK_colors_Teff.txt

particular metallicity value, but rather pool together the available evidence to report that LHS 1678 is metal poor.

### 3.2. Stellar Fundamental Parameters

Stellar mass, rotation, effective temperature, luminosity, and radius were estimated using multiple methods. Here we describe the method used to derive the values we adopt for all subsequent analyses. All other methods and values are described in Appendix A, with the exception of a

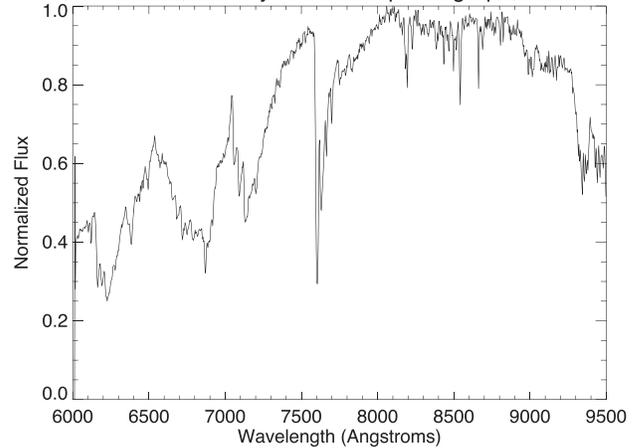

**Figure 1.** Normalized red-optical spectrum of LHS 1678 observed using RC Spec at the CTIO/SMARTS 1.5 m in 2005. The spectrum exhibits deep absorption bands from TiO, VO, and CaH from ∼6000–8000 Å, a characteristic of M dwarfs. We do not see the strong CaHn (n = 1–3) and TiO5 bands typically associated with cool subdwarfs (Gizis 1997). Following the methodology described in Henry et al. (2002), the spectrum yields a type M2.0V.

spectroscopically derived projected rotation velocity discussed in Section 4.3.

cTo estimate the stellar mass, we use the Benedict et al. (2016) mass–luminosity relations for absolute magnitudes $\mathcal{M}_V$ and $\mathcal{M}_K$ (their Equation (11)). These relations are calibrated using the individual masses of M dwarf binary components, rather than total system masses or model-based masses of single stars. For LHS 1678, we adopt the mean result of the $\mathcal{M}_V$ and $\mathcal{M}_K$ relations, $M = 0.345 \pm 0.014 \, M_\odot$.

Our methodology to derive effective temperature, luminosity, and radius is based upon that of Dieterich et al. (2014), to be presented in depth in M. L. Silverstein et al. (2022, in preparation). To derive the radius of 63 objects near the star/brown dwarf boundary, Dieterich et al. (2014) compared the BT-Settl 2011 photospheric models (Allard et al. 2012) to a variety of color combinations comprised of $(VRI)_B J H K_S W1 W2 W3$ photometry (subscript "B" for Bessell 1995 photometric system). Our procedure is an automated version of that code which uses $V_J R_{KC} I_{KC}$ instead of the Bessell system[43] and is applicable to the full range of M0V to L2.5V spectral types.

*Effective temperature.* By comparing an ensemble of observed colors to those extracted from the model grid we determine a best-fitting model with the smallest residuals for each of nine chosen colors. This provides an effective temperature ($T_{eff}$) for each color; we adopt the average as our result. $T_{eff}$ error is the standard deviation added in quadrature with 50 K, half the interval of the model grid. We note that we adopted a set model metallicity value of −0.5 for this procedure based on the photometric metallicity estimates previously described (see Section 3.1). Results from different stellar parameter estimation methods (see Appendix A) were also more consistent with each other when subsolar metallicity was assumed.

---
[43] Subscripts Johnson "J" and Kron-Cousins "KC" (or sometimes "C") for Johnson & Morgan (1953), Kron & Smith (1951), Cousins (1976) photometric systems, which come together as described by Landolt (2007).

**Table 2**
Host Star Derived Properties

| Property | Value | Error | Ref. |
|---|---|---|---|
| Mass ($M_\odot$) | 0.345 | 0.014 | Benedict et al. (2016)[a] |
| Rotation period (days) | 64 | 22 | Newton et al. (2017) Equation (6) |
| Effective temperature (K) | 3490 | 50 | This work[b] |
| Bolometric flux (log erg cm$^{-2}$s$^{-1}$) | −8.932 | 0.008 | This work[b] |
| Luminosity ($L_\odot$) | 0.0145 | 0.0003 | This work[b] |
| Radius ($R_\odot$) | 0.329 | 0.010 | This work[b] |
| Density (g cm$^{-3}$) | 13.624 | 1.40 | This work |
| Detected companions | 1 | ⋯ | This work (Section 3.5) |

**Notes.**
[a] Average of $\mathcal{M}_V$ and $\mathcal{M}_K$ relation values.
[b] Following the methodology of M. L. Silverstein et al. (2022, in preparation).

*Bolometric flux.* Once $T_{\rm eff}$ is derived, the model spectrum closest to the final value is iteratively scaled by a polynomial until observed and model photometry match to within 0.03 mag. The modified spectrum is then integrated within the wavelength range of the filters to get a partial flux. A bolometric correction is applied based on the amount of flux that would be missing from a blackbody of the same temperature viewed within our wavelength window. Bolometric flux error is derived using the mismatch between the final model and observed magnitudes and their observational errors.

*Bolometric luminosity, radius, and density.* The bolometric flux is scaled by the Gaia DR2 parallax to calculate luminosity, and radius is determined using the Stefan–Boltzmann law. Density is then calculated using our derived mass and radius. In a sample of 26 stars with interferometrically measured radii, the radii derived via this method matched those obtained via long-baseline interferometry to within an average of 6%. The estimated stellar parameter values used in this work are consolidated in Table 2.

### 3.3. Stellar Age

A combination of high galactic motion, low magnetic activity, low metallicity, and position on the HR diagram implies that this star is part of a population older than most of the thin disk.

*Galactic kinematics.* Combining our measured RV and Gaia DR2 astrometry, we have derived a total galactic motion with respect to the local standard of rest (LSR; Coşkunoğlu et al. 2011) of $94.6 \pm 0.4$ km s$^{-1}$ (see (UVW)$_{\rm LSR}$ in Table 1). This is a key property in identifying Galactic populations. Stars with total galactic motion between ∼85 and ∼180 km s$^{-1}$ are likely part of the Galactic thick disk (Nissen 2004; Bensby et al. 2014). Gan et al. (2020) use a method put forth by Bensby et al. (2003, 2014) to identify the TESS exoplanet system LHS 1815 as a thick disk member. Here we follow suit (according to Appendix A of Bensby et al. 2014), with less definitive results. $TD/T$ is the ratio of the probability of being in the thick disk to the probability of being in the thin disk. We use our (UVW)$_{\rm LSR}$ space motions to derive $TD/T = 0.44$. According to the metric of Bensby et al. (2014), this value indicates that LHS 1678 is likely in the thin disk, but just short of being classified an "in-between" star with galactic kinematics intermediate to the bulk of the thin and thick disk populations ($0.5 \leqslant TD/T \leqslant 2.0$). Following the metric adopted by Gan et al. (2020; $0.1 < TD/T < 10$), LHS 1678 is in between the two populations. We therefore conclude that the Galactic membership of LHS 1678 is uncertain, but the system is likely more kinematically heated and older than the average thin disk star.

*Low magnetic activity and rotation.* The TESS light curve spanning two sectors showed no signs of magnetic activity, either in the form of spot modulation or flares. We also extracted archival data from the All-Sky Automated Survey for Supernovae (ASAS-SN; Shappee et al. 2014; Kochanek et al. 2017; Jayasinghe et al. 2019) using the online tool[44] and the Gaia DR2 J2015.5 coordinates. The Lomb–Scargle (LS) periodograms of the 3.9 yr $V$- and 6.7 yr $g$-band data sets show no signs of rotation modulation (Figure 2), in line with our TESS findings, and only single-point outliers that cannot be confirmed as flares without higher cadence data. Our 16 yr CTIO/SMARTS 0.9 m data set, discussed in detail in Sections 3.5 and 3.6, also reveals low photometric variability and no flares and on longer timescales. Without spots in our time-series photometry, we are unable to derive a photometric rotation period. Our HARPS and CHIRON RV analyses indicate the star is a relatively slow rotator, with minimal line broadening (see Section 4.3), and we see H-$\alpha$ in absorption, which also points to a quiet, inactive star. We use the estimated mass of LHS 1678 and the mass–rotation relation of Newton et al. (2017; their Equation (6)) to estimate a stellar rotation period of $64 \pm 22$ days. We also estimate the rotation period via a $\log R'_{\rm HK}$–rotation relation. Using the HARPS data presented in Section 4.3, we derive a $\log R'_{\rm HK}$ value of $-6.087 \pm 0.548$ following the methods of Astudillo-Defru et al. (2017a), in agreement with the value of −6.08 from Rains et al. (2021). Substituting this value into the relations of Astudillo-Defru et al. (2017a) yields a rotation period ($P_{\rm rot}$) of $221 \pm 185$ days. This is slower than the vast majority of M dwarf rotation periods to date; we note that this relation is not calibrated to such a low value of $\log R'_{\rm HK}$ and this estimate is highly imprecise. We include it here, nonetheless, as part of our broad effort to estimate stellar rotation. We adopt the Newton et al. (2017) rotation period estimate of $64 \pm 22$ days and derive an age of 4–9 Gyr using the empirical relations of Engle & Guinan (2018) for M0–M1 dwarfs (yielding 3.9 Gyr) and M2.5–M6 dwarfs (yielding 8.5 Gyr). Because LHS 1678 is M2/M2.5 and the work of Engle & Guinan (2018) notes a lack of M1.5 to M2 dwarfs available for defining an M1.5–M2.0 relation, we use both available relations to estimate that the age of LHS 1678 is approximately 4–9 Gyr.

*Low metallicity, HR-diagram position, and spectral features.* Photometric metallicity estimates (Table 1, Section 3.1), better stellar parameter consensus with subsolar metallicity assumptions (Section 3.2, Appendix A), and low HR-diagram position (Figure 3) imply that the star has a metallicity less than zero. The lack of characteristically large CaHn ($n = 1$–3) and TiO5 band strengths in our RC Spec spectrum (Figure 1) suggests that LHS 1678 is not a cool subdwarf (Gizis 1997), in agreement with its position within the main sequence and its low tangential velocity (Jao et al. 2017). These findings are consistent with our assessments of galactic kinematics and low magnetic activity.

In aggregate, these properties indicate LHS 1678 is not a young star. As described, we estimate an age range of approximately 4–9 Gyr using rotation–age relationships. Other age considerations imply that LHS 1678 likely sits at the older end of this range.

---
[44] https://asas-sn.osu.edu/sky-patrol/coordinate/

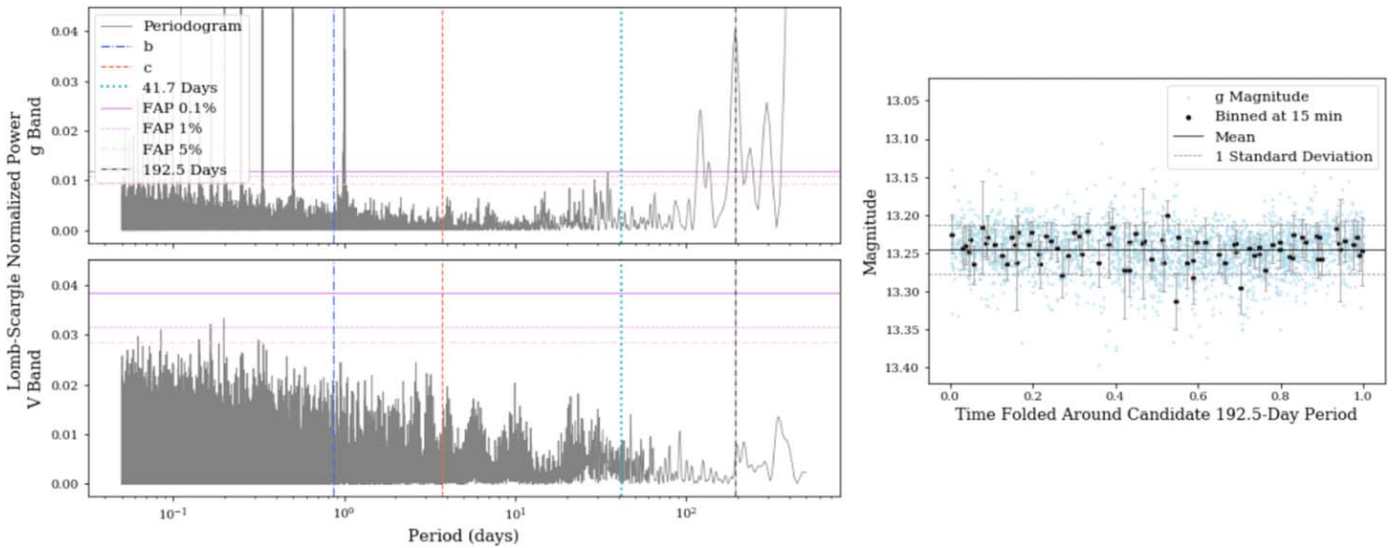

**Figure 2.** Nearly 10 yr of ASAS-SN data show no clear signs of stellar rotation. Left: `astropy`-generated LS periodograms of the LHS 1678 photometry. Horizontal lines denote false alarm probabilities (FAPs). Vertical lines denote the planet orbital periods, the 41.7 day HARPS signal to be discussed in Section 4.3, and the highest peak in the g-band LS periodogram at 192.5 days. Note that the peak in the g-band LS periodogram at one day is likely caused by observations being performed only at night in Chile and/or South Africa, and peaks at less than one day are likely associated aliases. We do not see any peaks in the V-band LS periodogram. Right: g-band ASAS-SN photometry folded on the candidate 192.5 day period found using the LS periodogram. Black circles are the data binned at a 15 minutes cadence, with error bars corresponding to the standard deviation of the magnitude values in each bin. We do not see clear signs of periodicity.

### 3.4. Unusual Position on the HR Diagram

We place LHS 1678 on the HR diagram to evaluate its position in the context of M dwarfs as a whole and relative to other M dwarf exoplanet host stars (Figure 3). With $\mathcal{M}_G = 10.047$ and $BP - RP = 2.278$ from Gaia DR2, the star occupies a narrow portion of $\mathcal{M}_G$ versus $BP - RP$ parameter space characterized by a gap in the lower main sequence. Jao et al. (2018) reported the gap in the HR diagram, revealed for the first time by high-precision parallaxes from Gaia DR2. More details on the gap and other features in the HR diagram are discussed by Jao & Feiden (2020) and Feiden et al. (2021), and the gap is successfully recovered in Gaia EDR3 (Figure 17 of Gaia Collaboration et al. 2021). The gap, or, more accurately described, a deficit of stars, is consistent with the expected transition from partially to fully convective interiors in M dwarfs. This gap is theoretically tied to the nonequilibrium burning and mixing of $^3$He in the stellar core. Core $^3$He fusion leads to the development of a convective core that grows and eventually merges with the convective outer envelope (van Saders & Pinsonneault 2012; Baraffe & Chabrier 2018; MacDonald & Gizis 2018; Feiden et al. 2021). Convective mixing then transports $^3$He away from the core, reducing the nuclear reaction rate and causing the core to contract and again separate from the convective envelope. This leads to the emergence of a $^3$He burning instability where the core undergoes damped, periodic transitions between partial and full convection until a balance is achieved in the $^3$He abundance and the star remains fully convective. This transitory phase in core energy transport leads to small, slowly varying oscillations in radius and luminosity in a narrow range of stellar masses spanning only ~0.34–0.37 $M_\odot$ (van Saders & Pinsonneault 2012;

Baraffe & Chabrier 2018) and manifests as the observed underdensity of stars in the HR diagram.

LHS 1678's location in, or at the lower edge of, the gap implies it may be part of a population of stars that have transitioned from being partially to fully convective, exhibiting long-term radial pulsations as they migrate back and forth across the gap over the course of billions of years. It is worth noting that this transition is theoretically predicted to be close to M3.5V, but spectral type is a poor marker compared to mass and age (Jao et al. 2018). Two independent analyses designate LHS 1678 as ~M2.0V (this work; Reid et al. 2007); at first glance, declaring this star as fully convective would seem to defy our current understanding of the transition region. It does not, however, because its observed photometry (Table 1) is consistent with the observed location of the gap and its mass, luminosity, and radius (Table 2) are all in agreement with multiple theoretical predictions of the parameter regime where M dwarfs undergo the $^3$He burning instability.

Figure 3 indicates that LHS 1678 is one of the only TESS M dwarf exoplanet systems currently known to reside in the gap. Only three other systems appear to be in or near the gap, and are reported as such for the first time here: GJ 357 is just above the gap at $BP - RP \approx 2.4$, and TOI-122 and LHS 1972 (GJ 3473) lie in the red end of the gap at $BP - RP \approx 2.6$ and 2.7, respectively. This redder part of the gap is more clearly illustrated in Figure D3 of Jao & Feiden (2020). The implications that gap membership has on planet formation and evolution are currently unknown. As more M dwarf planets are discovered and more precise stellar parameters become available (e.g., from Gaia DR3), trends in stellar parameters and exoplanet system properties in and near the gap should be monitored, with a focus on whether there are significant differences in the frequencies, distributions, and properties of planets. We discuss several potential impacts in Section 7.

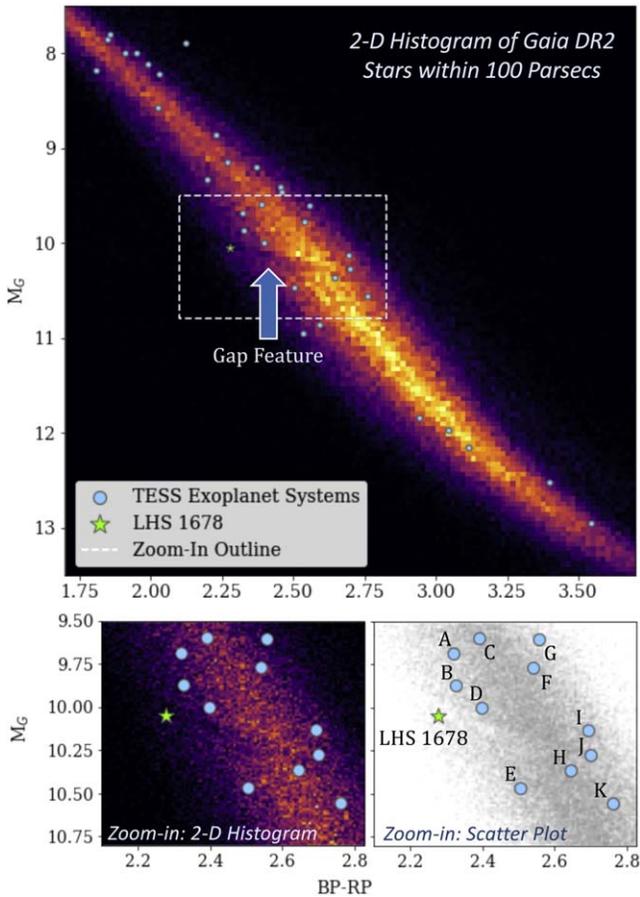

**Figure 3.** Gaia HR diagram depicting stars within 100 pc (using the "ABC" cut described by Lindegren et al. 2018), represented as a two-dimensional histogram (top, bottom left) and a scatter plot (gray points, bottom right). Overlaid are LHS 1678 (green star) and other TESS M dwarf exoplanet systems (blue circles). LHS 1678 is the only TESS exoplanet system so far that is associated with the observed gap in the lower main sequence. In the top panel, the gap is labeled with a blue arrow and is visually seen as the darker-colored, diagonal band, seemingly "cutting into" the main sequence from the left-hand side. The two bottom panels focus on the region around the gap and show LHS 1678's association with that feature. The next closest system, GJ 357 (labeled "D"), lies just above the gap at $BP - RP \approx 2.4$. TOI-122 ("H") and LHS 1972 ("J") are also close to or in a less visually obvious region of the gap at $BP - RP \approx 2.6$ $BP - RP \approx 2.7$. Left to right: TOI-1749, TOI-1235, TOI-1728, TOI-1899, TOI-532, LP 714-47, LP 961-53, L 168-9, AU Mic, TOI-1266, LHS 1815, TOI-1231, LHS 1678, GJ 1252 (labeled "A" in the bottom-right panel), TOI-270 ("B"), TOI-1201, TOI-700 ("C"), GJ 357 ("D"), TOI-1685, TOI-1634, L 98-59 ("E"), LHS 1478, TOI-674 ("F"), HATS-71 ("G"), LTT 1445A, TOI-122 ("H"), LTT 3780 ("I"), LHS 1972 ("J"), GJ 486 ("K"), TOI-237, TOI-2406, TOI-540, LHS 3844, LP 791-18 (see references in Appendix B).

### 3.5. Astrometric Detection of a Stellar or Substellar Companion

Sixteen years (2004–2020) of ground-based astrometry from the REsearch Consortium On Nearby Stars (RECONS) program at the CTIO/SMARTS 0.9 m telescope in Chile show compelling evidence that LHS 1678 hosts a low-mass stellar or substellar companion with an orbital period longer than the timespan defined by the data set. As initially reported by Jao et al. (2017), there is residual motion in the position of the LHS 1678 photocenter after solving for parallax and proper motion. Images of the field were acquired through the $V_J$ (henceforth simply $V$) filter at airmasses <1.1, meaning that corrections required for differential color refraction were minimal. Following the prescription described in Jao et al. (2005), the position of LHS 1678 was measured relative to five reference stars in 136 images taken on 30 nights with seeing better than 1″.8. Once a preliminary parallax and proper motion were measured, we subtracted the corresponding model from the data to get residuals in R.A. and decl. and fit an orbit to these residuals. Our methodology here is a least-squares approach to constraining orbital elements, as described in, e.g., Winters et al. (2017), guided by the work of Hartkopf et al. (1989). We adopt a set of starting values for the orbital period, epoch of periastron, and orbital eccentricity. We iteratively calculate an orbit and compare with the data, tweaking the starting values until we identify a best-fitting orbit. To derive our final parallax and proper motion, we subtracted the orbit model from the data and measured parallax and proper motion using the new data set. We derive a trigonometric parallax of $52.22 \pm 0.86$ mas and a proper motion of $992.3 \pm 0.2$ mas yr$^{-1}$ at position angle $166°.8 \pm 0°.1$. The high proper motion of the star allows us to rule out any contaminating background source that could affect the astrometry. For comparison, Gaia DR2 provides a parallax of $50.28 \pm 0.02$ mas and proper motion of $996.9 \pm 0.1$ mas yr$^{-1}$ at position angle $166°.1 \pm 0°.1$. For the majority of our analyses, we adopt Gaia astrometry. For the detailed astrometric characterization of LHS 1678 and its companion we use our RECONS values henceforth because they cover a much longer timespan than the Gaia measurements (16 yr versus 1.8 yr) and incorporate the higher-order motions due to the perturbation.

Figure 4 illustrates the photocenter positions, split into R.A. (top panel) and decl. (bottom panel) axes. Solid points represent the photocenter positions each night that multiple frames were taken (typically five frames over 10–20 minutes), where error bars represent standard deviations in the positions. Open points are from nights in which only single frames were taken or kept. The four points represented by "x" symbols on each axis are from observations between 2005 March and 2009 July taken through a different $V$ filter (because the original filter was cracked) than for the rest of the series. While the photometric results of this different filter match the original $V$ filter to a few percent, the astrometry is affected to varying degrees, particularly in the R.A. axis, depending on the configuration of the reference stars (see Subasavage et al. 2009). In the case of LHS 1678, the effects on the photocenter positions are relatively mild. These four points are used in the initial reduction for parallax and proper motion and not in the orbital fit.

An orbital fit traced with red curves in Figure 4 has been made to the photocentric positions on both axes, using the methodology of Hartkopf et al. (1989). The perturbation is evident only in decl., implying that the projected orbit of the companion is roughly north–south during the timespan of the observations. However, it is important to emphasize that the period resulting from the orbital fit, ~42 yr, is highly uncertain and likely underestimated because the current data do not cover enough of the orbit to constrain the orbital elements. Nonetheless, the current astrometric data imply that there is a companion of lower mass than the primary orbiting LHS 1678 with a period of at least several decades. This is long enough that it is unlikely to disturb the compact planetary system around the primary star unless the companion's orbit is highly eccentric; the primary star dominates the gravity well in the location of the planets (Kane 2019). We investigate the nature of this companion in depth in Section 7.3 and Appendix C and discuss the stability of the overall system further in Section 7.4.

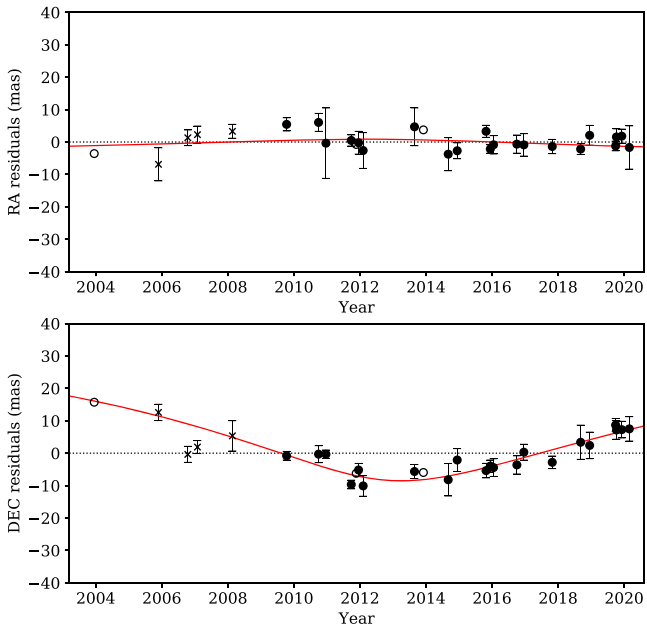

**Figure 4.** Residual astrometric motion of LHS 1678 when parallax and proper motion are accounted for and removed from 16 yr of monitoring. The remaining motion of the star indicates a binary system with an orbital period on the order of decades. The allowed periods and amplitude of the perturbation suggest a low-mass star or brown dwarf companion. The red line traces an orbital fit to the photocenter's motion. The four points marked with an "x" were taken in a slightly different V filter, and as such were not included in the orbital fit.

### 3.6. Long-term Variability

The images used for long-term astrometry can also be used to evaluate the photometric variability of LHS 1678 in the $V$ band. Each frame taken from 2004–2020 is represented by a point in Figure 5, where the flux of LHS 1678 has been compared to fluxes of five reference stars. The optical flux variations of the star are minimal, with a standard deviation from the average value of only 7.6 millimagnitudes; the high and low dotted lines represent the brightness levels at twice the standard deviation. This is, in fact, near the floor of measurements we make via the typical observational protocol at the CTIO/SMARTS 0.9 m (Hosey et al. 2015). A LS periodogram of these data also shows no signs of periodicity, in agreement with our ASAS-SN data findings in Section 3.3. We conclude that LHS 1678 shows a remarkably low level of magnetic activity, and exhibits no long-term photometric variations in $V$ above the 1% level during the 16 yr of observations. The lack of significant variability in these very-long-term observations is consistent with the absence of spot modulation and flares in the high-precision TESS observations covering a much shorter time baseline.

## 4. Follow-up Observations

### 4.1. Archival and High-resolution Imaging

The high proper motion of LHS 1678 allows us to inspect archival data, looking back in time when the star was elsewhere to see if there is another source at its sky position during the TESS observation epochs. The earliest archival image of the LHS 1678 field available on Aladin (Bonnarel et al. 2000)[45] is from the SuperCOSMOS Sky Survey (SCSS; Hambly et al. 2001). The image was taken in 1983 using the $B_j$ filter and a photographic plate, with a resolution of 0″.67 pixel$^{-1}$. Between that epoch and its late-2018 TESS observations, with its ≈1″ yr$^{-1}$ proper motion and 166° position angle, the star has moved about half an arcminute across the sky. This allows us to check for possible contaminating sources within the TESS photometric aperture and beyond, both by eye and using compiled catalog data (e.g., Gaia DR2), without fear of the star blocking a possible background source. The archival imaging revealed two faint sources within ≈1 TESS pixel of its location during Sectors 4 and 5.

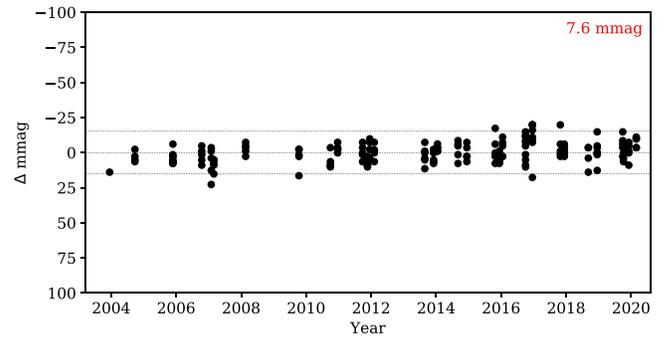

**Figure 5.** CTIO/SMARTS 0.9 m observations spanning ~16 yr reveal that LHS 1678 exhibits notably low magnetic activity. As a function of time, we examine the difference in $V_J$ magnitude relative to the mean magnitude of all observations (middle line). The standard deviation is 7.6 mmag; the top and bottom lines mark twice the standard deviation.

Figure 6 demonstrates the motion of LHS 1678 across the sky. 2′ × 2′ images from SCSS $B_j$ in 1983 (Figure 6(a)) and SCSS $I$ in 2000 (Figure 6(b)), taken directly from Aladin, reveal that there are no bright sources directly behind LHS 1678 during its TESS observations (Figure 6(c), produced using Lightkurve; Lightkurve Collaboration et al. 2018). All three images are centered on the Gaia DR2 J2015.5 coordinates of LHS 1678. The red dots in (c) are the brightest Gaia sources in the field of view. However, the aperture used in our TESS data analysis is very large (≈42″ in radius) and the change in flux from the transiting planets is very small. The TESS Sector 5 TPF of LHfg cvS 1678 in Figure 6(d) was generated using tpfplotter (Aller et al. 2020) and overlays the pipeline-defined extraction aperture used to create the LHS 1678 TESS light curves.[46] Critically, it includes two significantly fainter Gaia DR2 sources (red dots, #2 and #3). Both of these sources are fully blended within the TESS photometric aperture.

The source ~11″ to the NW of LHS 1678 (#2) is 8.751 magnitudes fainter than LHS 1678 in the $RP$ band, the closest Gaia band to the TESS band. If this faint source is a totally eclipsing binary, it would introduce a change in flux of about 0.016% the $RP$ flux of LHS 1678, which would be insufficient to reproduce either of the transit signals in this system, which correspond to flux changes of 0.037% and 0.074% for planets b and c, respectively.

---

[45] https://aladin.u-strasbg.fr
[46] This work made use of tpfplotter by J. Lillo-Box (Aller et al. 2020), publicly available at www.github.com/jlillo/tpfplotter The tpfplotter software makes use of the Python packages astropy, Lightkurve, matplotlib and numpy.

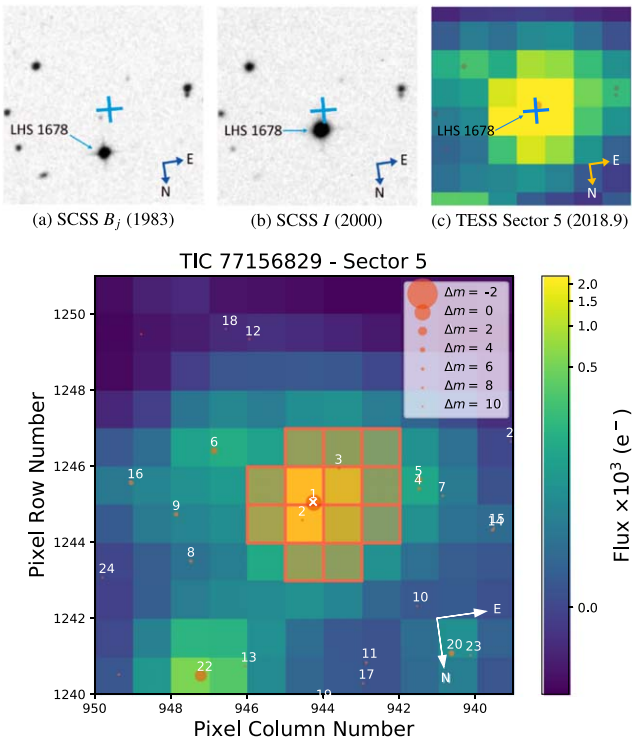

Figure 6. (a)–(c) Archival and TESS imaging of LHS 1678 and the surrounding sky demonstrating its motion across the sky from 1983 to late-2018 when TESS observations were acquired. Each image spans a $1.'5 \times 1.'5$ field of view approximately centered on the LHS 1678 Gaia DR2 coordinates, represented as a plus symbol. Made via Lightkurve (Lightkurve Collaboration et al. 2018) and labeled to match (a) and (b), (c) includes several bright Gaia DR2 sources as faint red dots to anchor the location of the closest bright sources in the earlier images. LHS 1678 dominates the flux detected by TESS, with no bright sources at its current location. There are, however, two faint sources nearby. (d) The TESS Sector 5 TPF image of LHS 1678 produced using the tpfplotter software package (Aller et al. 2020). The image spans $11 \times 11$ pixels with a pixel scale of $21''$ pixel$^{-1}$. The TESS SPOC pipeline photometric extraction aperture used to produce the Sector 5 light curve is shown as red shaded squares. Numbered red dots reveal the locations of Gaia DR2 sources with sizes scaled to their $\Delta G$ relative to LHS 1678. The faint sources #2 and #3 lie within the TESS aperture. We rule out these sources as the transit hosts via dilution arguments and follow-up observations (see Section 4.1).

The other source (#3) at $\sim 24''$ SE of LHS 1678 is 7.272 magnitudes fainter in *RP*. This corresponds to a factor of $\sim$780 in flux, a 0.062% change in the case that the faint source is a totally eclipsing binary, sufficient to masquerade as the planet b transit and potentially as planet c. We rule out these scenarios via ground-based time-series observations that recover the transits using photometric apertures that exclude this faint star (see Section 4.2). Neither of these sources are bright enough to significantly dilute LHS 1678's TESS light curve. Together they contribute only $\approx$0.155% of the flux in the TESS aperture.

These archival imaging analyses, at higher resolution than the TESS data, reveal there are no background sources at LHS 1678's current position that would significantly contaminate the light curve or mimic the observed transits. However, bound sources beyond the resolution limits of the archival imaging ($\sim 1''$) that are comoving with LHS 1678 could still remain. Such companions could be eclipsing binaries (EBs), host the transiting bodies, or add significant flux to the light curve and dilute the transit signals, leading to a bias in the derived planet

**Table 3**
Speckle Constraints

| Epoch | $\rho_{min}$ | Limiting $\Delta m$ | |
|---|---|---|---|
| ... | ... | $0.''15$ | $1.''0$ |
| ... | ($''$) | (mag) | (mag) |
| 2019.8594 | 0.0470 | 2.8 | 4.1 |
| 2019.9500 | 0.0415 | 2.4 | 4.9 |
| 2020.0182 | 0.0415 | 2.8 | 4.3 |

**Note.** $\rho_{min}$ indicates the resolution limit of the observation at the given epoch. Limiting $\Delta m$ values correspond to the *I*-band magnitude difference from LHS 1678 of a source that would be detectable at the given separation (i.e., $0.''15$ and $1.''0$). No sources were detected.

radii (Ciardi et al. 2015; Furlan & Howell 2017). To identify close-in bound companions, we observed LHS 1678 via both speckle imaging using the Southern Astrophysical Research (SOAR) High Resolution Camera (HRCam)+SOAR Adaptive Optics Module (SAM) setup and adaptive optics using the Very Large Telescope (VLT) NAOS-CONICA instrument (NaCo).

LHS 1678 was observed three times with the HRCam on the SAM (Tokovinin et al. 2016) at the SOAR 4.3 m telescope. Optical speckle data were collected on UT 2019 November 10, UT 2019 December 13, and UT 2020 January 7 in the *I* band ($\lambda_c = 824/170$ nm), which is approximately centered on the TESS bandpass (Table 3). The observations used integrations of 6–25 milliseconds and were acquired in blocks of 400 images. The data were processed following the procedures described in detail in Tokovinin (2018) and Ziegler et al. (2020) to produce two-dimensional speckle autocorrelation "images" and contrast sensitivity curves. The HRCam data revealed LHS 1678 to be a single star within the limits of the observations. The speckle imaging limits are presented in Table 3. Their implications for the nature of the astrometric companion described in Section 3 are described in Section C.2.

We also observed LHS 1678 using the NaCo instrument (Rousset et al. 2003; Lenzen et al. 2003) on the 8.2 m VLT Unit Telescope 1 (UT1) on UT 2019 August 17 to obtain adaptive-optics-corrected near-IR images. We acquired nine images in the $Br\gamma$ filter ($\lambda_c = 2.166$ μm), each with an integration time of 20 s. The frames were dithered between observations to use the science data for sky subtraction in postprocessing. We used custom software to perform bad-pixel correction, flat fielding, sky-background subtraction, image registration, and coadding. The final reduced image (Figure 7, inset) had a pixel scale of $0.''013221$ pixel$^{-1}$. We estimated our contrast sensitivity (Figure 7) by injecting simulated companion sources at discrete separations from the host star and scaling their luminosities until they were recovered at $5\sigma$. The reduced data do not reveal any close-in companions within a few arcseconds of LHS 1678. Details on the limits placed on bound companions by the NaCo observations and their implications for the nature of the astrometric companion are presented in Section C.2.

### 4.2. Ground-based Time-series Photometry

To redetect the transit signals attributed to the TESS planet candidates, refine transit depths and ephemerides, and rule out nearby sources as contaminating nearby EBs, we pursued seeing-limited, ground-based time-series observations. These observations were organized, obtained, and analyzed through TESS Follow Up Observing Program (TFOP)[47] Subgroup 1 (SG1). The observations were scheduled using the TESS Transit Finder, a modified version of the Tapir software (Jensen 2013). A summary of the observations obtained through SG1 for each candidate planet transiting LHS 1678 is provided in Table 4. This includes observation dates and parameters and key analysis results for each ground-based time series.

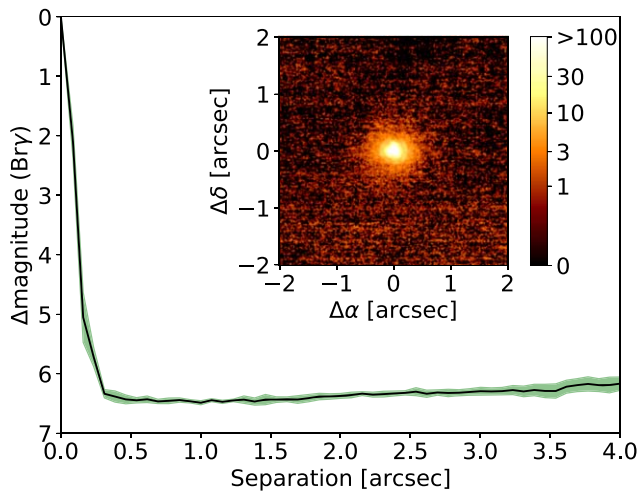

**Figure 7.** VLT NaCo adaptive-optics image of LHS 1678 (inset) and corresponding 5σ contrast sensitivity curve with a 1σ error bar. Flux in the inset image is in log scale, ranging from 0 to 100 detector units, with the core of the stellar point-spread function (PSF) saturated. No companions are detected down to the contrast limits, including the astrometrically detected companion described in Section 3.

LCOGT—We obtained ten transit observations of the inner planet candidate (TOI-696.01) and three transit observations of the outer planet candidate (TOI-696.02) using the 1.0 m telescopes at three southern hemisphere sites of the Las Cumbres Observatory Global Telescope Network (LCOGT; Brown et al. 2013). Each telescope is equipped with a SINISTRO CCD camera with a pixel scale of $0.39''$ pixel$^{-1}$ and $26'$ field of view. Observations were performed in either the $zs$ or $I_c$ filters. The images in each time series were calibrated using the LCOGT Banzai pipeline (McCully et al. 2018). The AstroImageJ software package (Collins et al. 2017) was used to extract differential photometry from the images, produce light curves, and investigate nearby EBs (within $2.5'$) at the periods of the planet candidates. The aggregated results of these LCOGT observations and analyses were that six transits of the inner planet candidate were detected at the anticipated times and with the expected depth (although with a low signal-to-noise ratio (S/N)), one partial transit of the outer planet candidate was detected, and the possibility of a nearby EB causing the apparent TESS transits was ruled out at the periods of both planet candidates.

PEST—A single transit of the inner planet candidate was observed using the 0.3 m Perth Exoplanet Survey Telescope (PEST).[48] PEST carries an SBIG ST-8XME CCD camera with a plate scale of $1.2''$ pixel$^{-1}$ and a $31' \times 21'$ field of view. The observations were performed in the $R_c$ filter. Image reduction, differential photometry, and light curve analyses were performed using the custom C and Python-based PEST pipeline.[49] No transit-like signal was detected at the predicted time and with the expected depth of TOI-696.01, and the data were insufficient to clear nearby EBs.

MEarth—We used seven of the eight 0.4 m telescopes in the MEarth-South telescope array (Nutzman & Charbonneau 2008; Irwin et al. 2015) at CTIO in Chile to perform defocused observations of a transit of the outer planet candidate. The MEarth-South telescopes carry Apogee U/F230 CCD cameras that produce images with a pixel scale of $0.84''$ pixel$^{-1}$ and $29' \times 29'$ field of view. The observations were performed in a custom broad red-optical filter (*RG715*). Image reduction and aperture photometry was performed using custom software, resulting in seven individual light curves of LHS 1678 across the predicted transit time. In the combined light curve, a significant transit is detected at the predicted time and with the anticipated transit depth.

Here we detail the five LHS 1678 b observations that were excluded from our analysis. *PEST 2019-07-30*: The apparent event was during a time with some passing clouds, so is likely systematics driven. *LCOGT 2019-08-06*: The observation was not defocused, so one of the potential comparison (comp) stars was saturated, and thus not useful. The resulting best light curve scatter was too high to rule out or confirm the event. Event detection is further complicated by very little preingress baseline data. *LCOGT 2020-03-14*: The apparent late and too deep event seems to be driven by including a specific comp star in the comparison ensemble. Without the problematic comp star included, the scatter in the data is too high to confirm or rule out the transit. *LCOGT 2020-08-03*: This light curve shows a possible on-time 400 ppm event, but there is a mid-transit upswing in the tentative detection, so we elected to exclude it from the joint model. *LCOGT 2020-11-10*: Changes in full-width half-maximum (FWHM) and sky transparency at the time of expected ingress likely caused systematics that confuse any shallow transit detection. The data neither rule out or confirm the shallow event. LHS 1678 c ground-based transits that were excluded from our analysis had similar issues with observing conditions.

The ground-based time-series data indicate that the signals detected in TESS photometry are consistent with planets orbiting LHS 1678 and rule out nearby stars as sources of EB contaminants. The ground-based light curves presented here were mostly gathered before it was discovered that the outer planet is in a 3.7 day orbit, rather than the initially identified 14.8 day period (see Section 2). The exception is the 2020 March 16 MEarth light curve, which was obtained just before the star's sky position moved too far toward the Sun to be observed. These data capture a transit at half the initially identified 14.8 day period. More recent, ongoing monitoring to search for transit timing variations (TTVs), to be presented in another paper, are consistent with the outer planet having a 3.7 day orbit. We incorporate the ground-based transit detections reported here into an analysis of the exoplanet properties in Section 5. The ground-based data included in the joint analysis are marked in Table 4 and the light curves with the joint model fits are shown in Figures 12 and 13. We also include the constraints imposed by these data in our statistical validation of the planet candidates in Section 6.

---

[47] https://tess.mit.edu/followup/
[48] http://pestobservatory.com
[49] http://pestobservatory.com/pipeline-overview

**Table 4**
Ground-based Time-series Observations[a]

| Telescope | Epoch (UT) | Camera | Filter | Pixel Scale (″ pixel$^{-1}$) | PSF[b] Size (″) | Aperture Radius (pixels) | Duration (min.) | # of Obs. | In Fit?[d] (Y/N) | Notes |
|---|---|---|---|---|---|---|---|---|---|---|
| ... | | ... | ... | | | | | | | ... |
| LHS 1678 b | | | | | | | | | | |
| LCO-CTIO-1.0 m | 2019-07-27 | Sinistro | $zs$ | 0.39 | 2.32 | 20 | 124 | 163 | Y | Possible detection |
| PEST-0.3 m | 2019-07-30 | ST-8XME | $R_c$ | 1.23 | 4.6 | 6 | 180 | 72 | N | Passing clouds, inconclusive |
| LCO-SSO-1.0 m | 2019-08-06 | Sinistro | $zs$ | 0.39 | 1.55 | 15 | 117 | 100 | N | No detection? |
| LCO-CTIO-1.0 m | 2019-12-05 | Sinistro | $I_c$ | 0.39 | 4.01 | 25 | 185 | 200 | Y | Detection |
| LCO-SAAO-1.0 m | 2019-12-24 | Sinistro | $zs$ | 0.39 | 3.2 | 14 | 192 | 133 | Y | Detection |
| LCO-SAAO-1.0 m | 2020-03-14 | Sinistro | $zs$ | 0.39 | 3.5 | 16 | 140 | 91 | N | Late[c], too deep, high airmass, inconclusive |
| LCO-SAAO-1.0 m | 2020-07-26 | Sinistro | $zs$ | 0.39 | 5.21 | 17 | 136 | 58 | Y | Detection |
| LCO-SSO-1.0 m | 2020-08-03 | Sinistro | $zs$ | 0.39 | 5.3 | 17 | 162 | 102 | N | No detection? |
| LCO-SAAO-1.0 m | 2020-08-20 | Sinistro | $zs$ | 0.39 | 3.17 | 17 | 213 | 135 | Y | Detection |
| LCO-SAAO-1.0 m | 2020-09-28 | Sinistro | $zs$ | 0.39 | 4.62 | 16 | 231 | 146 | Y | Detection |
| LCO-SSO-1.0 m | 2020-11-10 | Sinistro | $zs$ | 0.39 | 4.33 | 20 | 235 | 147 | N | No detection? |
| LHS 1678 c | | | | | | | | | | |
| LCO-SSO-1.0 m | 2019-08-14 | Sinistro | $zs$ | 0.39 | 2.26 | 15 | 178 | 161 | N | Late[c] |
| LCO-CTIO-1.0 m | 2019-09-13 | Sinistro | $zs$ | 0.39 | 1.86 | 16 | 186 | 160 | Y | Detection |
| LCO-CTIO-1.0 m | 2020-02-08 | Sinistro | $zs$ | 0.39 | 2.96 | 20 | 243 | 197 | N | No detection? |
| MEarth-South-x7-0.4 m | 2020-03-16 | Apogee U/F230 | $RG715$ | 0.84 | 4.4 | 9.9 | 239 | 1170 | Y | Detection |

**Notes.**
[a] All ground-based transit observations were observed as continuous time series.
[b] This column indicates which light curves were (Y) and were not (N) used in the modeling described in Section 5.
[c] "Late" denotes that the **event** was detected later than the time predicted by the ephimerides.
[d] Point-spread Function

### 4.3. Spectroscopy

To check for false positives in the form of a spectroscopic binary or background EB not resolved by the previously described follow-up, we acquired eight RV measurements using the CHIRON echelle spectrograph (Schwab et al. 2010; Tokovinin et al. 2013; Paredes et al. 2021) on the CTIO/SMARTS 1.5 m in Chile (program ID 19A-0339; see Table 8). Our measurements, spanning 410–870 nm, were taken from 2019 August 27 to September 6 using slicer mode with $R \sim 80{,}000$. RVs were calculated assuming a $v \sin i$ of $0.53 \text{ km s}^{-1}$ using the same methodology that Winters et al. (2020) applied to their TRES data.[50] As described in A. A. Medina et al. (2022, in review), we measure the $v \sin i$ as follows. We search for the maximum peak correlation via two nested-grid searches. The first searches over a $v \sin i$ range of 0-100 km s$^{-1}$, sampled at 1 km s$^{-1}$ intervals, while the second samples the best value from the first search within a range of $\pm 1$ km s$^{-1}$ at 0.1 km s$^{-1}$ intervals. We then use parabolic interpolation to obtain the final $v \sin i$ value of 0.53 km s$^{-1}$ from the second grid search. This value is within the $v \sin i$ upper limit of 1.8 km s$^{-1}$ set by half the spectral resolution of CHIRON. The top panel of Figure 8 shows as an example the 7100 Å TiO bands from the 2019 August 27 spectrum and the cross-correlation function (CCF) that yields our RV measurement for that date and set of bands. RVs were derived using all six available TiO band apertures in the spectrum and by taking the weighted mean, with the exception of 2019 September 2, for which we only used the one with the highest S/N ("aperture 44"). Errors of 100 m s$^{-1}$ are adopted for each measurement except that of 2019 September 2, which had a low S/N and for which we used only one TiO band, and 2019 September 6, which had a higher dispersion across different TiO band measurements. As demonstrated in Table 6, we do not detect double lines indicative of a spectroscopic binary in the individual spectra, nor any variation in RVs across the full CHIRON observation baseline that would be characteristic of a wider orbit EB or brown dwarf across a range of periods. This result is not inconsistent with the low-mass companion detected astrometrically (Section 3.5), because that companion's motion in 2019 was likely more tangential than radial (see Figure 4). See Appendix C for additional discussion on this point.

Following the TESS detection of the two candidate planets, LHS 1678 was included in a precision RV monitoring program (ESO Programme ID: 1102.C-0339) on the HARPS echelle spectrograph (Pepe et al. 2002; Mayor et al. 2003) on the ESO 3.6 m telescope at La Silla Observatory, Chile. Forty-one RV observations were collected between UT 2019 November 15 and UT 2020 March 18 with exposure times of 1800 s at a spectral resolution of 115,000. To mitigate contamination in the blue end of the stellar spectrum, the spectra were observed without the on-sky calibration fiber.

High-precision RVs were measured from HARPS spectroscopic data using the maximum-likelihood template fitting method described in Astudillo-Defru et al. (2017b). In this method, a high-S/N stellar template spectrum is constructed by shifting all observed spectra of the target star to a common RV reference frame and coadding the shifted spectra. A combined telluric spectrum is also computed by subtracting the stellar template from each observed spectrum and coadding the residuals in the rest frame. The RV for each observed spectrum is determined by finding the velocity offset that maximizes the likelihood between the spectrum and the stellar template. The telluric template is included in this process to mask out spectral regions most affected by sky contamination. The individual HARPS RVs measured using this process are provided in

---

[50] https://github.com/mdwarfgeek/tres-tools

---

**Table 5**
RV Data

| Epoch (BJD) | RV (+11,300 m s$^{-1}$) | Error (m s$^{-1}$) | Instrument |
|---|---|---|---|
| 2458723.8503 | 92 | 100 | C |
| 2458729.8728 | 69[a] | …[a] | C |
| 2458724.8783 | 83 | 100 | C |
| 2458726.9320 | 87 | 100 | C |
| 2458727.9116 | 111 | 100 | C |
| 2458728.9120 | 90 | 100 | C |
| 2458729.8730 | 148 | 100 | C |
| 2458730.8925 | 106 | 100 | C |
| 2458733.9152 | 57[b] | …[b] | C |
| 2458802.624889 | 169.30 | 1.27 | H |
| 2458803.593582 | 170.16 | 2.21 | H |
| 2458804.853906 | 168.42 | 1.83 | H |
| 2458805.652423 | 168.68 | 1.63 | H |
| 2458806.638706 | 166.81 | 1.85 | H |
| 2458806.786704 | 168.46 | 2.00 | H |
| 2458807.824849 | 170.10 | 1.86 | H |
| 2458820.750258 | 164.99 | 1.50 | H |
| 2458824.817474 | 165.96 | 1.98 | H |
| 2458836.644053 | 162.97 | 2.05 | H |
| 2458837.554549 | 164.31 | 2.12 | H |
| 2458838.795811 | 166.49 | 3.68 | H |
| 2458839.672360 | 168.46 | 2.12 | H |
| 2458848.691204 | 165.95 | 1.35 | H |
| 2458850.777971 | 170.11 | 1.84 | H |
| 2458851.701479 | 168.06 | 1.42 | H |
| 2458852.543461 | 167.97 | 2.04 | H |
| 2458854.707755 | 167.60 | 2.92 | H |
| 2458866.730163 | 161.98 | 2.34 | H |
| 2458868.562685 | 165.95 | 1.62 | H |
| 2458876.652162 | 164.74 | 4.37 | H |
| 2458880.569614 | 168.20 | 2.43 | H |
| 2458882.638426 | 165.40 | 1.81 | H |
| 2458884.545161 | 172.03 | 3.27 | H |
| 2458885.692906 | 169.35 | 2.97 | H |
| 2458888.663975 | 167.38 | 2.23 | H |
| 2458889.531907 | 167.19 | 1.95 | H |
| 2458890.590408 | 167.53 | 2.21 | H |
| 2458894.659895 | 167.43 | 2.04 | H |
| 2458897.656124 | 162.37 | 2.49 | H |
| 2458898.593149 | 165.04 | 1.70 | H |
| 2458900.624242 | 168.03 | 2.52 | H |
| 2458901.579623 | 169.53 | 1.87 | H |
| 2458902.521406 | 165.91 | 1.45 | H |
| 2458910.546586 | 165.26 | 1.66 | H |
| 2458911.520510 | 164.10 | 1.65 | H |
| 2458914.509491 | 166.24 | 1.86 | H |
| 2458916.576156 | 161.86 | 2.03 | H |
| 2458917.508347 | 162.73 | 1.55 | H |
| 2458924.506351 | 171.09 | 1.61 | H |
| 2458926.507216 | 168.75 | 1.54 | H |

**Notes.** H: HARPS; C: CHIRON. CHIRON errors taken to be 100 m s$^{-1}$, with $v \sin i = 0.53$ km s$^{-1}$ assumed.
[a] Value is from only one spectrum's order 44 (TiO bands at 7100 Å), rather than the usual two spectra and six orders.
[b] Low signal-to-noise ratio.

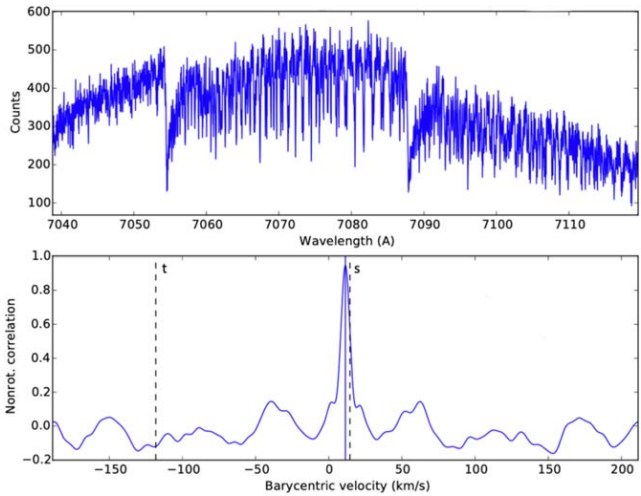

**Figure 8.** A cross-correlation analysis focusing on strong TiO absorption bands in the CHIRON spectra of LHS 1678 reveals no evidence for double lines (top) and a single-peaked cross-correlation function (bottom). These analyses rule out very close-in, unresolved binary companions. The stability of LHS 1678's RV measured by CHIRON over the full observing baseline further excludes the possibility of a spectroscopic companion orbiting across a range of masses and periods (Table 6). The vertical, dashed lines labeled $s$ and $t$ in the cross-correlation plot indicate where the sky and telluric lines would be; our template does not correlate with those lines.

Table 8. The scatter in the HARPS RV measurements was 2.5 m s$^{-1}$ with a median measurement uncertainty of 2 m s$^{-1}$.

To explore sources of systematic noise in the RV time series, we use standard outputs from the HARPS Data Reduction Pipeline (DRS; Lovis & Mayor 2007) and calculate stellar activity indicators for each spectrum. The DRS computes radial velocities and other diagnostics from each observed spectrum using the CCF technique. Two diagnostics that are produced are the FWHM of the CCF and the RV bi-sector span (BIS). We also measure several stellar activity indicators following the methods described in Astudillo-Defru et al. (2017b). These include the S-Index, multiple lines in the hydrogen Balmer series (H$\alpha$, H$\beta$, H$\gamma$), and the sodium D doublet (NaD). We show LS periodograms of the RV time series and each additional diagnostic in Figure 9. These are computed using the `LombScargle` algorithm in the Python package `astropy`. We calculated 1% and 5% false alarm probabilities (FAPs) for each time series using a bootstrap approach.

The LS periodogram of the RV data exhibits no significant peaks (above 1% FAP), including at the known periods of the planetary signals. Several other diagnostics exhibit significant peaks at ~1 day, consistent with the observation cadence. The peak with the highest power in the RV periodogram is at ≈42 days. We further investigated this signal and identified a coherent trend in the RV data at 41.7 days (Figure 10). This candidate signal is intriguing and could represent the stellar rotation period or a harmonic. Stellar rotation estimates of LHS 1678 (Section 3.3) are in broad agreement with this signal. The LS periodograms of several activity indicators exhibit structure at similar periods, although none are significant (Figure 9). Alternatively, the signal could be suggestive of an additional longer-period planet in the system. Such nontransiting planets have been identified in many M dwarf systems (e.g., K2-18, GJ 357; Cloutier et al. 2019a; Luque et al. 2019). We fit a Keplerian orbit model with broad input priors to further investigate this hypothesis, shown in Figure 10. The data can be represented by a Keplerian orbit model at a period of 41.7 days with some level of eccentricity, although it is poorly constrained ($e = 0.2$–0.6). The nature of this candidate signal remains unconstrained by the available data. Continued RV monitoring and observations with higher precision are warranted to investigate further and measure the planet masses (see Section 7).

We used our CHIRON and HARPS measurements to compute the weighted-mean RV of LHS 1678. The combined HARPS and CHIRON RV is $11.4667 \pm 0.0259$ km s$^{-1}$ (HARPS only: $11.4669 \pm 0.0259$, Kervella et al. 2017; CHIRON only: $11.39 \pm 0.50$ km s$^{-1}$). The absence of RV variation with time allows us to exclude the possibility of a grazing transit by nonplanetary mass companion as a false positive in our transit detections. In Section 5, we jointly combine the TESS and ground-based transit observations and HARPS RV measurements to place constraints on the planet properties, including mass upper limits.

## 5. Planet Parameter Estimation

We inferred the properties of the two planets using data from the TESS photometric time-series, ground-based transits observed by MEarth and LCOGT, radial velocities from HARPS, and the stellar parameters described in Section 3. Our analysis method follows a very similar procedure to that described in Kostov et al. (2019a) and Gilbert et al. (2020), with the addition of RV data and ground-based transits and some minor parameterization changes in the model. We obtained SPOC PDCSAP (Stumpe et al. 2012, 2014; Smith et al. 2012) instrumental systematics-corrected light curves from the Mikulski Archive for Space Telescopes (MAST) using the Python package `Lightkurve` (Lightkurve Collaboration et al. 2018). We note that the PDCSAP light curves from Sectors 4 and 5 used in our analysis were subject to small overestimates of the sky-background flux due to a bias in the original algorithm.[51] Fortunately, LHS 1678 is relatively uncrowded, and the transit depth in these sectors was biased to be ~1% deeper, while the radius was biased to be ~0.5% larger, both an order of magnitude smaller than our error bars. To estimate the planet properties, we jointly modeled the exoplanet transits from TESS and ground-based data, stellar variability in the light curves, along with other systematics and the HARPS RV data using the `exoplanet` package (Foreman-Mackey et al. 2020). All data sets were modeled simultaneously to preserve covariances between parameters. The ground-based photometry included in the model is listed in Table 4 with a "Y" in the "In Fit?" column.

Each TESS sector is modeled with a mean offset and a white noise variance term described using a log-normal distribution. We modeled residual stellar variability using a Gaussian process (GP) model (Foreman-Mackey et al. 2017; Foreman-Mackey 2018) that describes a stochastically driven damped harmonic oscillator with two hyperparameters, power ($S_0$) and angular frequency ($\omega_0$). The ground-based light curves also include a parameter to model additional variance in the data that is not encompassed in the reported observational uncertainty, and the same GP model as the TESS data, but with independent hyperparameters. Each observatory has a

---

[51] See DRN 38 for Sector 27 for more information on the sky-background bias. Note that the sky-background algorithm was corrected for sectors forward from Sector 27 and that the reprocessed data for Sectors 1–13 are not subject to the sky-background bias.

Table 6
CHIRON Constraints

| Mass | 10 day Period | | | 20 day Period | | | 200 day Period | | |
|---|---|---|---|---|---|---|---|---|---|
| | K | Δ RV | Det. | K | Δ RV | Det. | K | Δ RV | Det. |
| Equal mass | 38 | ... | N | 30 | ... | N | 14 | ... | N |
| 0.075 $M_\odot$ | 13 | 5 | N | 10 | 2 | N | 4.8 | 0.1 | N |
| 10 $M_J$ | 1.9 | 0.760 | N | 1.5 | 0.300 | N | 0.7 | 0.014 | ? |

**Note.** In the case of an equal-mass companion the spectrum would appear with double lines, which were not detected; we note this, rather than include Δ RV in that case. Mass: companion mass; K: RV semiamplitude (km s$^{-1}$); Δ RV: change in RV (km s$^{-1}$ day$^{-1}$); Det.: detected ($N$ = no, ? = insufficient sensitivity).

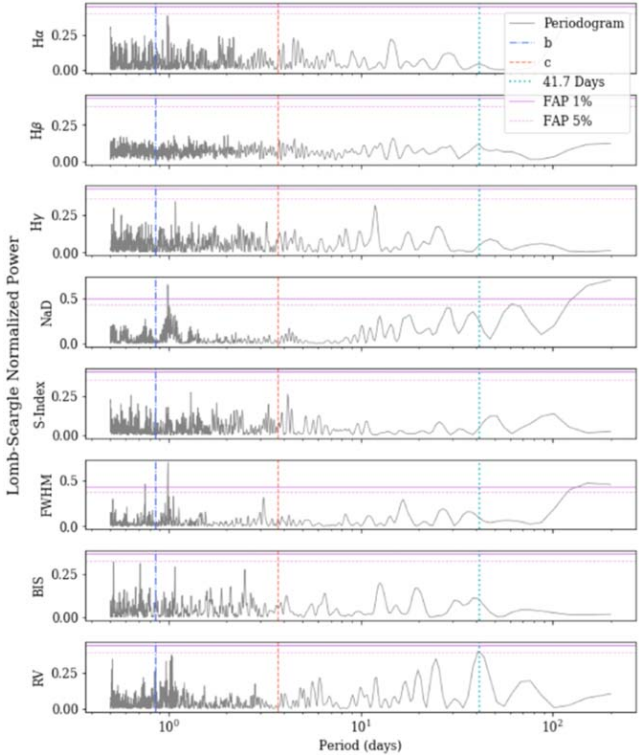

Figure 9. LS periodograms of HARPS RVs, activity indicators, and other diagnostics. The 5% and 1% FAPs are shown as solid and dashed horizontal lines. The periods of the TESS candidate planets derived from our joint modeling (Section 5) are plotted as vertical dotted–dashed (planet b) and dashed (planet c) lines. The RV periodogram exhibits no significant periodic signals at those orbital periods. The strongest signal in the RV data, with a 5% FAP, appears at 41.7 days (vertical dotted line). The signal may be associated with the stellar rotation period or a harmonic, but the data are insufficient to place a firm constraint. This candidate signal does not appear with any significance in the activity diagnostics. We note that several diagnostics exhibit significant peaks at ∼1 day, consistent with the observation cadence.

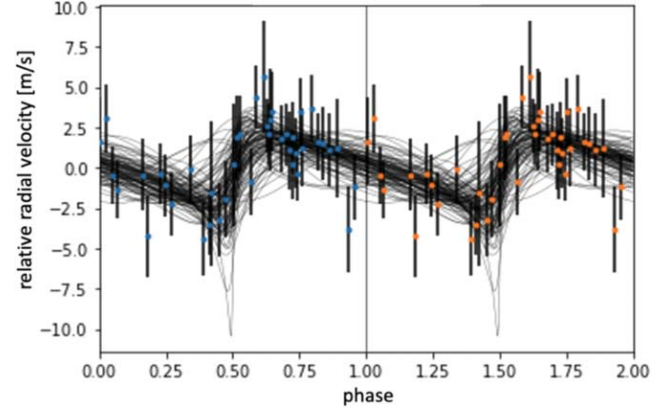

Figure 10. HARPS RV data folded on the candidate signal at 41.7 days, repeated across two full periods, with no binning, to better show the periodic nature. This signal may be consistent with the rotation period of LHS 1678 or a harmonic. Alternatively, the signal may be evidence of a longer-period planet in the system. To explore this hypothesis, we fit the signal with a Keplerian orbit model assuming broad input priors. We show random draws from the fit posterior as black curves. While the data can be fit with such a model, the true nature of the candidate signal at 41.7 days remains unconstrained with the available data.

separate model parameter for the additional uncertainty that is added in quadrature and included in the likelihood function. The model also included the stellar and planetary parameters. The stellar inputs included the stellar density ($\rho_*$), the stellar radius ($R_*$), and quadratic limb-darkening parameters ($u$) parameterized following Kipping (2013) with different limb darkening for the various bandpasses. For each planet, the model included the midpoint time of first transit ($T_0$), impact parameter ($b$), ratio of planet radius to star radius ($R_p/R_s$), orbital eccentricity ($e$), longitude of periastron ($\omega$), and planet mass. Instead of the standard method of parameterizing the exoplanet orbits as a function of period, we opted to parameterize the model in terms of the midpoint time of the last transit, making the orbital period of a planet a deterministic parameter in the model, rather than a parameter we sample in. This parameterization yielded more efficient sampling when including the ground-based photometric data. The choice of prior probability distributions for each of these modeled parameters was the same as described in Gilbert et al. (2020) except we used a normal distribution for the time of the last transit. We also included a log-normal prior on planet mass with a wide standard deviation of 3 $M_\oplus$. For the RV model, we included a quadratic trend and a jitter term.

We used PyMC3 (Salvatier et al. 2016) to perform a Hamiltonian Monte Carlo sampling from the posterior model. We drew over 100,000 independent samples, a very large number, to ensure that we were able to obtain well-sampled upper limits for the planet mass distributions. The central 68th percentile of folded transit models for the TESS data are shown in Figure 11 and for the ground-based transit observations in Figures 12 and 13. The ground-based transit observations are also stacked in Figure 14 with the model overlaid to better demonstrate the successful recovery of both transits from the ground. As expected, incorporation of ground-based observations, acquired after TESS Sectors 4 and 5, yielded better constraints on the model than in the case where only TESS data were used. In particular, the derived radius of the outer planet increased slightly. The median model parameters and 1σ uncertainties computed during the PyMC3 sampling are provided in Table 7. We measured radii of $0.696 \pm 0.044$ and $0.982^{+0.064}_{-0.063}$ $R_\oplus$ for the inner and outer planets, respectively.

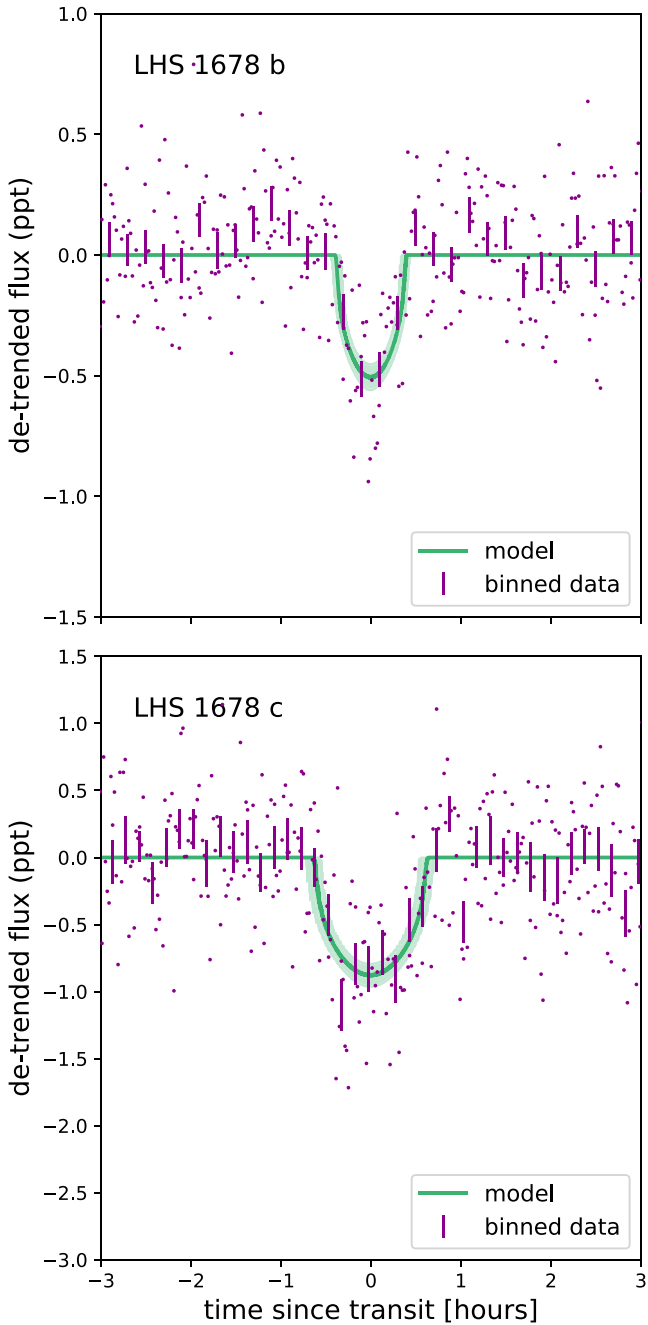

Figure 11. TESS photometry folded on the planet periods. Purple dots are the individual 2 minute cadence measurements, and the vertical lines are 10 minute bins. The best-fit transit models described in the text are overlaid as green curves. Green shading corresponds to the central 68th percentile (1σ) range of models consistent with the data.

Figure 15 shows the full HARPS RV time-series and best-fit model and uncertainty for RV signal and trend. No significant RV signal from the planets were detected, but we were able to compute 97.7th percentile mass upper limits for each planet, 0.35 $M_\oplus$ for the inner, and 1.4 $M_\oplus$ for the outer.[52] Additional parameters are listed in the bottom portion of Table 7. Figure 16 shows the posterior mass distributions and 2 and 3σ limits from the joint modeling. We adopt the 2σ upper limits because, for planets of these sizes, the 3σ values correspond to densities that are unphysical (see Section 7.2 and Figure 17).

### 6. Validation

The detection of multiple TESS transits combined with ground-based follow-up and precision RV observations provides a compelling data set that indicates there is little parameter space remaining for false positives that may mimic the transit signals. To place a numerical value on the remaining FPP and statistically validate the planets, we use the vespa tool (Morton 2015). vespa combines the host star properties, the observed TESS transits, and follow-up constraints and compares to simulated false-positive scenarios allowed by the remaining parameter space in a probabilistic framework. The output of vespa is the likelihood that the detected transit signal may be mimicked by the simulated false-positive scenarios, the FPP. We ran the algorithm on each planet candidate individually. The required inputs included the host star position, broadband photometry, and stellar parameters (see Tables 1 and 2). The depth of the second-deepest, eclipse-like feature in the phase-folded light curve is also included as a required input. We also included additional constraints on background EBs from our ground-based time-series observations[53] and constraints on bound companions from archival imaging, high-resolution imaging, and RV monitoring. We find FPPs of $1.02 \times 10^{-5}$ for LHS 1678 b and $1.70 \times 10^{-5}$ for LHS 1678 c. These FPPs are low enough (≪1%) to consider the planets statistically validated. In addition to the low statistical probabilities of remaining false positives, we note that our follow-up time-series observations effectively rule out all nearby stars as background EBs, and our CHIRON and HARPS RV monitoring rules out bound companions across a wide range of periods and masses. Additionally, false positives are less likely in multiplanet systems (Lissauer et al. 2012), and the FPPs estimated here could be considered upper limits. Analysis of TESS multiplanet systems indicates they may be further reduced by ∼50× by this multiplicity boost (Guerrero et al. 2021).

These considerations of false positive scenarios do not take into account the wide-orbit, astrometrically detected stellar or substellar companion. The astrometric monitoring data indicate that the companion is most likely a brown dwarf (see Section 7.3). Assuming that the companion is an early-L dwarf ≈8 mag fainter than LHS 1678 in the $I_c$ band,[54] not even a totally eclipsing binary brown dwarf could reproduce the transit depths of either planet in the similar TESS band, given the significant dilution from the primary. Furthermore, it would only account for one of the transit signals associated with LHS 1678, as a multiply eclipsing brown dwarf system at the

---

[52] Note that the distribution used here is actually log mass, which we convert to mass for our listed values. The 2σ value is calculated at the more massive end of the distribution, corresponding to the 97.7th percentile.
[53] This is in the form of the minimum separation for EBs ruled out by ground-based time-series observations, or the maximum size of the apertures used for ground-based light curve extraction: 9.8″ for LHS 1678 b and 8.3″ for LHS 1678 c (see Table 4).
[54] Estimated using the color-temperature relation table at http://www.pas.rochester.edu/~emamajek/EEM_dwarf_UBVIJHK_colors_Teff.txt.

Table 7
Planet Parameters

| Parameter | Median | $+1\sigma$ | $-1\sigma$ |
|---|---|---|---|
| *Measured Parameters* | | | |
| **Star** | | | |
| $\rho$ [g cm$^{-3}$] | 13.9 | 1.5 | 1.5 |
| Limb darkening TESS $u_1$ | 0.96 | 0.35 | 0.45 |
| Limb darkening TESS $u_2$ | −0.22 | 0.47 | 0.32 |
| Limb darkening LCO $u_1$ | 0.60 | 0.53 | 0.42 |
| Limb darkening LCO $u_2$ | 0.03 | 0.42 | 0.41 |
| Limb darkening MEarth $u_1$ | 0.95 | 0.56 | 0.61 |
| Limb darkening MEarth $u_2$ | −0.21 | 0.54 | 0.43 |
| **LHS 1678 b** | | | |
| First transit ($T_0$) [BJD—2457000] | 1411.476805455 | 0.00101 | 0.00094 |
| Last transit [BJD—2457000] | 2121.168814342 | 0.00522 | 0.00385 |
| Period [days] | 0.8602322 | 0.0000068 | 0.0000048 |
| Impact parameter[a] | 0.24 | 0.16 | 0.16 |
| $R_p/R_*$ | 0.0192 | 0.0010 | 0.0010 |
| Radius [$R_\oplus$] | 0.696 | 0.044 | 0.044 |
| Mass [$M_\oplus$][†] | 0.06 | 0.10 | 0.05 |
| Eccentricity[a] | 0.041 | 0.066 | 0.032 |
| $\omega$ [radians][a] | 0.1 | 2.2 | 2.3 |
| $a/R_*$ | 8.16 | 0.29 | 0.30 |
| $a$ [AU] | 0.01251 | 0.00059 | 0.00056 |
| Inclination [deg] | 87.55 | 1.60 | 1.43 |
| Duration [hours] | 0.774 | 0.068 | 0.085 |
| Insolation [$S_\oplus$] | 93.2 | 9.3 | 8.4 |
| **LHS 1678 c** | | | |
| First transit $T_0$ [BJD—2457000] | 1414.759411086 | 0.00149 | 0.00147 |
| Last transit [BJD—2457000] | 2120.360616833 | 0.00399 | 0.00310 |
| Period [days] | 3.694247 | 0.0000024 | 0.0000021 |
| Impact parameter[a] | 0.39 | 0.14 | 0.21 |
| $R_p/R_*$ | 0.0272 | 0.0015 | 0.0016 |
| Radius [$R_\oplus$] | 0.982 | 0.064 | 0.063 |
| Mass [$M_\oplus$][a] | 0.39 | 0.43 | 0.27 |
| Eccentricity[a] | 0.034 | 0.070 | 0.029 |
| $\omega$ [radians][a] | −0.01 | 2.1 | 2.1 |
| $a/R_*$ | 21.56 | 0.76 | 0.79 |
| $a$ [AU] | 0.0331 | 0.0016 | 0.0015 |
| Inclination [deg] | 89.09 | 0.57 | 0.46 |
| Duration [hours] | 1.26 | 0.10 | 0.12 |
| Insolation [$S_\oplus$] | 13.5 | 1.3 | 1.2 |
| *Upper Limits* | | | |
| | $1\sigma$ | $2\sigma$ | $3\sigma$ |
| **LHS 1678 b** | | | |
| Eccentricity | 0.11 | 0.25 | 0.46 |
| Mass [$M_\oplus$] | 0.16 | 0.35 | 0.76 |
| Density [g cm$^{-3}$] | 2.7 | 6.0 | 12.8 |
| Est. RV Semiamplitude [m s$^{-1}$] | 0.2 | 0.5 | 1.0 |
| **LHS 1678 c** | | | |
| Eccentricity | 0.11 | 0.22 | 0.37 |
| Mass [$M_\oplus$] | 0.82 | 1.4 | 2.1 |
| Density [g cm$^{-3}$] | 6.0 | 10.6 | 17.0 |
| Est. RV Semiamplitude [m s$^{-1}$] | 0.7 | 1.2 | 1.8 |

**Note.**
[a] These values are poorly constrained, but are included for complete disclosure of the model output and our uncertainty considerations.

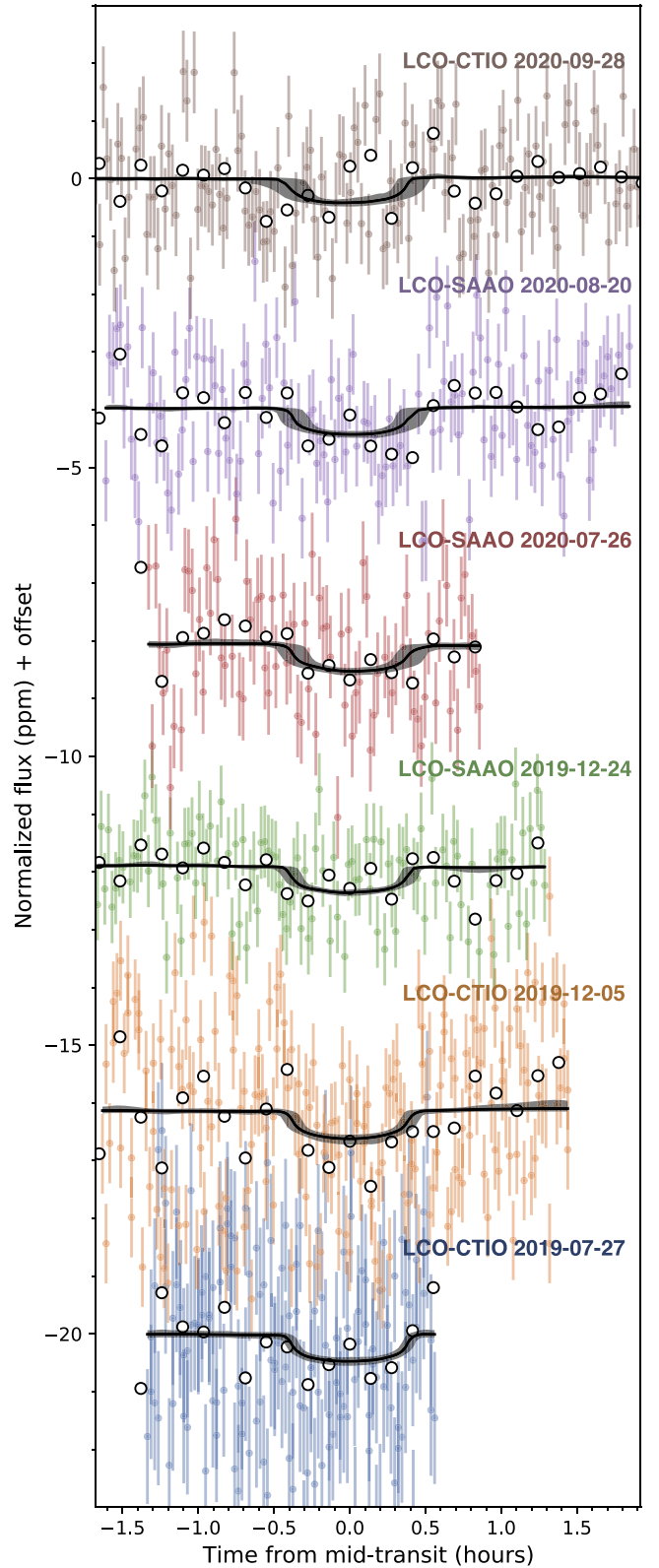

**Figure 12.** Ground-based transit observations of LHS 1678 b. Data points with uncertainties represent individual photometric measurements in each time series and white circles are data binned at 10 minutes cadence. The best-fit model is represented as a black line with $1\sigma$ uncertainties in gray shading. We note that the uncertainty in the transit times increases with time since the TESS observations, as expected.

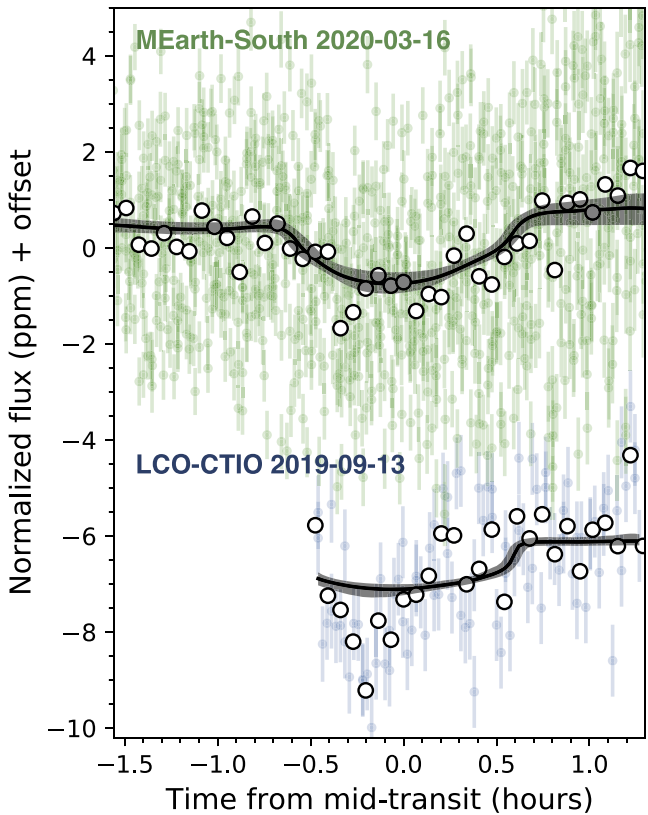

**Figure 13.** Ground-based transit observations of LHS 1678 c. The data points and curves are derived the same way as in Figure 12, but the binned data (white circles) are at 5 minutes cadence.

observed periods would likely be unstable. If the wide companion is instead a very-low-mass star, the observed signals could be caused by a heavily diluted transiting giant planet, but such a system would be exceptionally rare and, again, only account for one of the observed signals.

## 7. Discussion

The LHS 1678 system is comprised of an older-population, magnetically quiescent M2 dwarf at 19.9 parsecs, with a wide-orbit, low-mass star or brown dwarf companion, and two validated, roughly Earth-sized (0.7 $R_\oplus$, 1 $R_\oplus$) planets in 0.9 day and 3.7 day orbits around the primary star. Via HARPS radial velocity measurements, we derive 97.7th percentile mass upper limits of 0.35 $M_\oplus$ and 1.4 $M_\oplus$ for planets b and c, corresponding to density upper limits of 6.0 g cm$^{-3}$ and 10.6 g cm$^{-3}$, respectively. The planet radii and mass upper limits, when compared to similarly sized planets (Table 8) and planet composition models (Marcus et al. 2010; Zeng et al. 2019), indicate they are likely to be rocky (Figure 17).

Although the presence of at least two small planets orbiting an M dwarf in a compact configuration is now known to be commonplace, several properties of the LHS 1678 system mark it as exciting in the context of current cutting-edge studies and observations. The inner planet is an ultra-short-period planet (USP; Winn et al. 2018), with a sub-day orbital period of ∼21 hr. It is a compelling candidate for JWST observations to study its atmosphere via thermal emission. The outer planet is in the Venus zone (VZ; Kane et al. 2014), receiving ∼13.5× the flux of Earth, and a possible data point in studying the runaway greenhouse effect in Earth-sized planets. The relatively old age of the star presents an additional constraint that provides useful insight into planetary evolutionary processes and possible periods of past temperate surface conditions (Way et al. 2016; Kane et al. 2019; Way & Del Genio 2020). The system is prime to search for evidence of additional planets via further transit searches and TTVs; in Section 7.1 we discuss a third planet candidate. Both of the known planets probe models of planet formation and are prime candidates for further mass-constraining precision RV observations (Section 7.2). This system has a faint low-mass star or brown dwarf companion with an orbital period on the order of decades (Section 7.3). The system is dynamically stable such that there could be additional nontransiting planets with periods intermediate to the two known (Section 7.4). The host star is associated with the gap in the HR diagram lower main sequence tied to the M dwarf convective boundary; physical mechanisms associated with this HR-diagram position have unknown effects on exoplanet formation and evolution (Section 7.5).

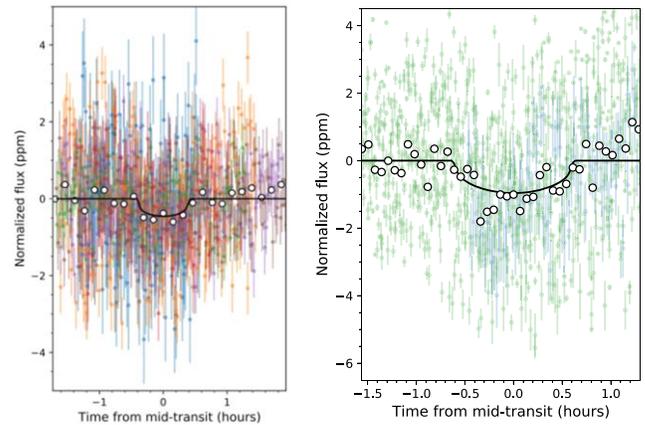

**Figure 14.** To demonstrate more clearly the ground-based detections of each planet, the ground-based light curves for planets b (left) and c (right; see Table 4) are stacked on top of each other and binned at 5 minutes cadence (white circles), with the transit model (black curve) overlaid.

### 7.1. Search for Additional Planets

After we completed the main analysis of the TESS and follow-up data presented here, LHS 1678 was observed for two additional sectors (UT 2020 October 21 to December 17) in Cycle 3 of the TESS extended mission, as part of Guest Investigator Programs G03228,[55] G03272,[56] G03274,[57] and G03278,[58] the Cool Dwarf target catalog (Muirhead et al. 2018), and the TESS CTL (Stassun et al. 2018b). Because at the time of writing, the complete two-cycle, four-sector set of TESS 2 minute cadence measurements for this star had not yet been searched by the SPOC pipeline, we performed our own search of this combined data set to see if any additional planet signals would be revealed. Two independent analyses were performed and are described here.

*Analysis 1.* We calculated a box least-squares (BLS; Kovács et al. 2002) periodogram, as implemented and optimized by

---

[55] High Frequency Quasi-Periodic Pulsations In M Dwarf Flares – PI: C. Million.
[56] Two Minute TESS Data For Thousands Of Promising New Radial Velocity Target Stars – PI: J. Burt.
[57] Understanding The Physical Origin Of The Rocky/Non-Rocky Transition Around Mid-To-Late M Dwarfs With TESS – PI: R. Cloutier.
[58] Enriching Our View Of Multiplanet Systems Using TESS – PI: A. Mayo.

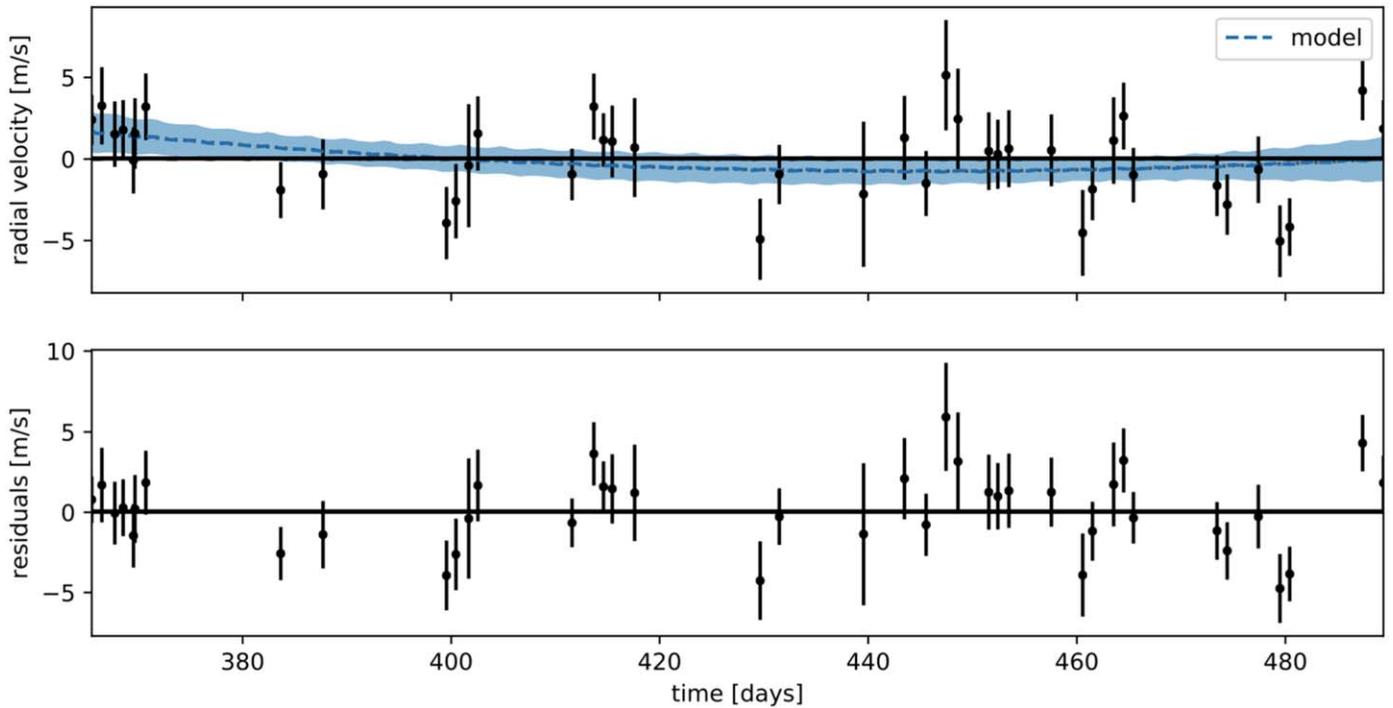

**Figure 15.** Top: full HARPS RV time series with individual measurement uncertainties (black) along with the best-fit model and 1$\sigma$ uncertainty (blue dashed line and shading). No significant RV signal is detected at the period of either planet nor is a significant RV trend. Bottom: RV residuals after removing the best-fit model.

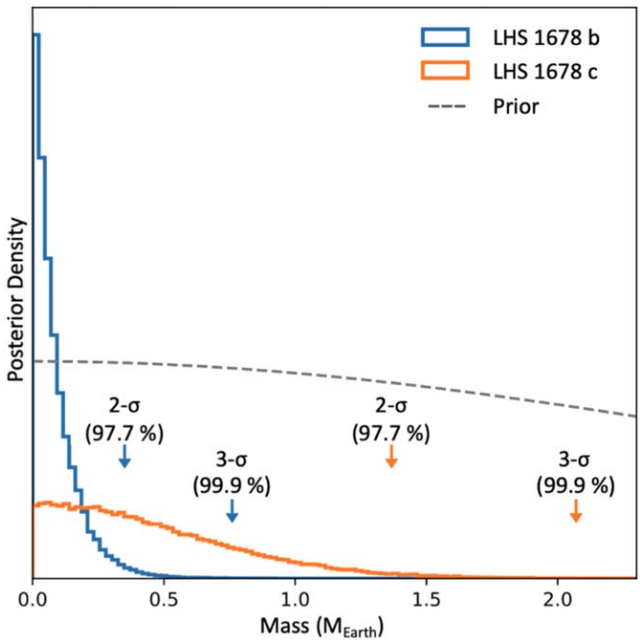

**Figure 16.** Posterior mass distributions for each planet resulting from our HARPS RV data modeling. We place 2$\sigma$ and 3$\sigma$ upper limits on the mass of each planet (corresponding to the usual 97.7% and 99.9% values, respectively), shown as color-coded arrows. The prior probability distribution is represented by a dashed gray line. The 3$\sigma$ values correspond to bulk planet densities greater than iron, so we adopt the 2$\sigma$ values as our upper limits.

Vanderburg et al. (2016) on the full Cycle 1 (Sectors 4 and 5) and Cycle 3 (Sectors 31 and 32) 2 minute data sets. Our search confirmed the detection of the two planets found by the SPOC pipeline presented earlier in this work, and also detected an additional sub-Earth-sized transiting planet candidate with a 4.965 day orbital period. The newly detected candidate has a relatively low S/N of about 9.4 in the BLS search, so we investigated the robustness of this new signal by searching for it in the TESS full frame image (FFI) data.

We downloaded FFIs in the region around LHS 1678 using the TESScut interface (Brasseur et al. 2019) and extracted our own light curve. We corrected for systematics using a different method than the SPOC light curve, namely by decorrelating against the quaternion time series following Vanderburg et al. (2019), and still recovered the same 4.965 day signal, indicating that it is unlikely to be due to an instrumental systematic. We also extracted a light curve from a single-pixel aperture centered on the position of LHS 1678 and detected the 4.965 day transit signal as well, implying that the signal must originate within about 20″ of the position of the target star. Henceforth, we refer to the new candidate as TOI-696.03.

We then performed a Markov Chain Monte Carlo (MCMC) fit to a multicycle SPOC PDCSAP 2 minute cadence light curve to estimate TOI-696.03's parameters. We removed points taken during transits of the other two planets in the system and modeled the light curve with Mandel & Agol's (2002) transit model (Figure 18, top panel). We assumed that the candidate has a circular orbit and imposed a prior on the stellar density presented in Table 2. We also imposed priors on the quadratic limb-darkening coefficients of $u_1 = 0.17 \pm 0.15$ and $u_2 = 0.44 \pm 0.15$ based on models from Claret & Bloemen (2011), and sampled the coefficients using the $q_1$ and $q_2$ parameterization of Kipping (2013). We explored the parameter space using the differential evolution MCMC algorithm of Ter Braak (2006), evolving 100 chains for 20,000 links, and discarding the first 1000 links to remove the burn-in phase. The MCMC yielded a period of $4.965222 \pm 0.000023$ days, a radius of $0.91^{+0.08}_{-0.08}\,R_\oplus$, a time of transit of BJD = 2458806.8143 ± 0.0015, and a scaled semimajor axis (a/R$_*$) of 26.1 ± 0.9. The period-folded transit along with the best-fit transit model from this analysis is presented in the top panel of Figure 18.

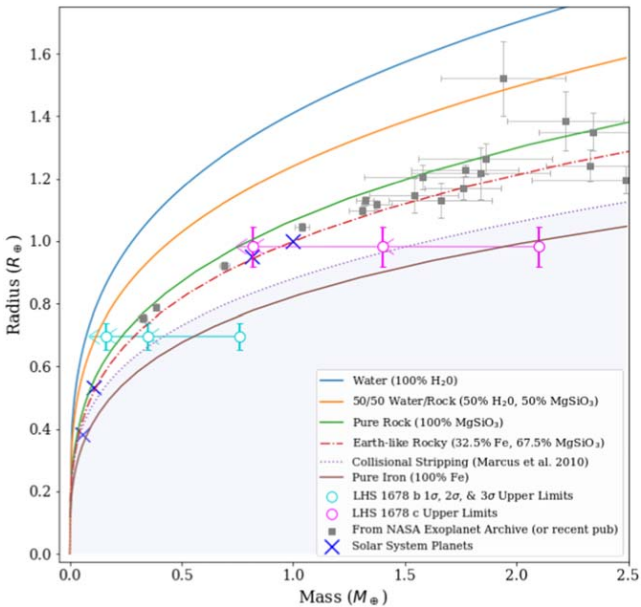

**Figure 17.** Mass–radius diagram with LHS 1678 b and c, the rocky solar system planets, and small exoplanets with robust measurements of $M < 2.5\,M_\oplus$, $R < 1.6\,R_\oplus$, and errors of <20%. In order of increasing mass these are TRAPPIST-1 h, d, e, f, c, g, and b, LTT 1445 A c, L 231-32 b, GJ 1132 b, LHS 1140 c, Kepler-78 b, GJ 357 b, GJ 3473 b, L 98-59 d, L 98-59 c, LHS 1478 b, LTT 3780 b, K2-229 b, and LTT 1445 A b (properties and references in Table 8). LHS 1678 b and c are each labeled with three open circles with arrows connecting them from right to left to denote the 3-, 2-, and 1σ (99%, 97.7%, and 68%) mass upper limits. Lines of constant density for a variety of compositions are overlaid (Zeng et al. 2019). Planets with densities that lie beneath the collisional stripping curve (Marcus et al. 2010; Zeng et al. 2019) should be nonphysical, so that region is shaded. The collisional stripping mass upper limits of LHS 1678 b and c are 0.419 $M_\oplus$ and 1.51 $M_\oplus$, respectively. Notably, the majority of planets on this diagram are of similar density to an Earth-like rocky planet. This, combined with our mass constraints, indicates that the LHS 1678 planets are likely to be predominantly rocky. Of particular note, the 1σ mass upper limit for LHS 1678 c, which is in the "Venus zone", places it identically with Venus on the mass–radius diagram within our error bars.

*Analysis 2.* Independent of Analysis 1, we also performed a separate BLS search on the SPOC PDCSAP light curves from TESS Cycles 1 and 3, also following Kovács et al. (2002) and Vanderburg et al. (2016), and recovered LHS 1678 b's and c's signals at the periods reported earlier in this work. After masking out both signals, we found a third transit signal with a period of 4.96519 d and transit epoch of 2458414.564464 BJD at an S/N of 9.5. This corresponds to a predicted radius of ∼0.9 Earth radii and transit duration of 1.024 hr.

The signal is also recovered in the `Quick-Look Pipeline` (QLP; Huang et al. 2020a, 2020b) FFI light curves following a similar process, although with somewhat lower significance. This is likely due to the relatively short duration of TOI-696.03, which compromises the transit depth when undersampled with the longer FFI cadences.

To verify that the transit signal originates from the target star, we compute difference images from the TESS FFIs (Figure 19) and the publicly available Python package `TESS-plots`.[59] First, `TESS-plots` uses `TESSCut` to create a pixel cutout of the FFIs centered on the target star. Then, each transit's in- and out-of-transit cadences are identified using the signal's ephemerides, and the means of the in- and out-of-transit frames are calculated. The mean out-of-transit image represents the target star with no transits, while the difference between out-of-transit and in-transit images reveals the location of the largest source of variability during a transit. We excluded all cadences that occurred during the transit of other planets before calculating the difference images to ensure that any source of variability was indeed due to the newly identified planet candidate, TOI-696.03.

---

[59] https://github.com/mkunimoto/TESS-plots

We inspected the difference images for all three signals across Cycles 1 and 3. Figure 19 shows our results for TOI-693.03 in the most recently observed sector (Sector 32). As shown in Figure 19, the location of the strongest difference for transits of TOI-696.03 coincides with the pixel of the target star, as is expected in the case of a transiting planet orbiting LHS 1678. The dots in the images show the location of nearby TIC stars, down to $\Delta T = 4$ mag. No nearby stars within this magnitude range lie in the same pixel as the target star, indicating that the planet candidate is likely to be on-target. For the other planets in the system, LHS 1678 b and LHS 1678 c, our independent difference image result is consistent with the SPOC difference image analysis.

We fit the transit of TOI-696.03 to derive its parameters using the `exoplanet` toolkit. The stellar mass and radius were fit with a normally distributed prior using the values in Table 2. We used a quadratic limb-darkening model, and both period and epoch had uniform priors around the values we found from our BLS search. The resulting parameters and best-fit model are shown in Table 9 and the bottom panel of Figure 18.

*A likely third planet near-resonance with LHS 1678 c.* Two independent analyses of the full Cycle 1 and Cycle 3 TESS 2 minute and FFI data recover an additional planet candidate, TOI-696.03. This candidate has a period of 4.965 days and a radius of ≈0.9 $R_\oplus$. With currently available data, we find no reason to suspect that this signal is either an instrumental or astrophysical false positive. The transit signal is most likely consistent with the presence of another planet in the LHS 1678 system. Because we identified this additional candidate late in our analyses of the system, we were not able to collect ground-based transit follow-up observations and do not attempt to formally validate this relatively low-S/N signal here. We do note that soon after these analyses were performed, SPOC released a multisector, multicycle search result that identified planet c at the correct 3.7 day period and revealed a planet candidate, with properties matching the one independently identified here (e.g., 4.97 day period, $R_p \approx 1\,R_\oplus$). Ground-based time-series follow up of this candidate required for validation is ongoing using TFOP resources.

Both analyses and the SPOC multicycle search identify TOI-696.03 at a period that is within 1% of 4:3 mean-motion resonance with LHS 1678 c. In Kepler planetary systems with three or more planets, there are nine pairs of planets within 1% of 4:3 resonance; all of the pairs are in resonance chains with at least one other planet in the system (notably the Kepler-223 Mills et al. 2016 and Kepler-60 Steffen et al. 2013 systems). This has been interpreted as a signature of convergent disk migration (e.g., Goldreich & Schlichting 2014; Tamayo et al. 2017). However, LHS 1678 b is far away from any resonances

**Table 8**
Planets with Robust Mass Measurements and $R < 1.6\,R_\oplus$, $M < 2.5\,M_\oplus$ as of 2021 Oct 20

| System | Dist.[1] (pc) | $K_s^*$ (mag) | SpTy | $T_\mathrm{eff}$ (K) | Mass ($M_\odot$) | Ref. | Planet | Period (days) | Radius ($R_\oplus$) | Semimajor Axis (au) | Insol. Flux ($S_\oplus$) | Mass ($M_\oplus$) | Disc. Ref. | Prop. Ref. |
|---|---|---|---|---|---|---|---|---|---|---|---|---|---|---|
| TRAPPIST-1 | 12.4 | 10.296 ± 0.023 | M8.0V ± 0.5 | 2566 ± 26 | 0.0898 ± 0.023 | G00, C06, A21, M19 | b | $1.510826^{+0.000006}_{-0.000006}$ | $1.116^{+0.014}_{-0.012}$ | $0.01154^{+0.00010}_{-0.00010}$ | $4.153^{+0.161}_{-0.159}$ | $1.374^{+0.069}_{-0.069}$ | G17 | A21 |
| | | | | | | | c | $2.421937^{+0.000018}_{-0.000018}$ | $1.097^{+0.014}_{-0.012}$ | $0.01580^{+0.00013}_{-0.00013}$ | $2.214^{+0.086}_{-0.085}$ | $1.308^{+0.056}_{-0.056}$ | | |
| | | | | | | | d | $4.049219^{+0.000026}_{-0.000026}$ | $0.788^{+0.011}_{-0.010}$ | $0.02227^{+0.00019}_{-0.00019}$ | $1.115^{+0.043}_{-0.043}$ | $0.388^{+0.012}_{-0.012}$ | | |
| | | | | | | | e | $6.101013^{+0.000035}_{-0.000035}$ | $0.920^{+0.013}_{-0.012}$ | $0.02925^{+0.00025}_{-0.00025}$ | $0.646^{+0.025}_{-0.025}$ | $0.692^{+0.022}_{-0.022}$ | | |
| | | | | | | | f | $9.207540^{+0.000032}_{-0.000032}$ | $1.045^{+0.013}_{-0.012}$ | $0.03849^{+0.00033}_{-0.00033}$ | $0.373^{+0.015}_{-0.014}$ | $1.039^{+0.031}_{-0.031}$ | | |
| | | | | | | | g | $12.352446^{+0.000054}_{-0.000054}$ | $1.129^{+0.015}_{-0.013}$ | $0.04683^{+0.00040}_{-0.00040}$ | $0.252^{+0.010}_{-0.010}$ | $1.321^{+0.038}_{-0.038}$ | | |
| | | | | | | | h | $18.772866^{+0.000214}_{-0.000214}$ | $0.755^{+0.014}_{-0.014}$ | $0.06189^{+0.00053}_{-0.00053}$ | $0.144^{+0.006}_{-0.006}$ | $0.326^{+0.020}_{-0.020}$ | | |
| LTT 1445 A | 6.9 | 6.496 ± 0.021 | M2.5V | 3337 ± 150 | 0.257 ± 0.014 | D15, W21 | c | $3.1239035^{+0.0000034}_{-0.0000034}$ | $1.147^{+0.055}_{-0.054}$ | $0.02661^{+0.00047}_{-0.00049}$ | $10.95^{+2.35}_{-1.98}$ | $1.54^{+0.20}_{-0.19}$ | W21 | W21 |
| | | | | | | | b | $5.3587657^{+0.0000043}_{-0.0000042}$ | $1.305^{+0.066}_{-0.061}$ | $0.03813^{+0.00068}_{-0.00070}$ | $5.36^{+1.18}_{-0.96}$ | $2.87^{+0.26}_{-0.25}$ | W19 | W21 |
| L 231-32 | 22.5 | 8.251 ± 0.029 | M3.5V ± 0.5 | 3506 ± 70 | 0.40 ± 0.02 | P13, G19, VE21 | b | $3.3601538^{+0.0000048}_{-0.0000048}$ | $1.206^{+0.039}_{-0.039}$ | $0.03197^{+0.00022}_{-0.00022}$ | $19.0^{+0.3}_{-0.3}$ [†1] | $1.58^{+0.26}_{-0.26}$ | G19 | VE21 |
| GJ 1132 | 12.6 | 8.322 ± 0.027 | M4.5V | 3270 ± 140 | 0.181 ± 0.019 | BT15 | b | $1.628931^{+0.000027}_{-0.000027}$ | $1.130^{+0.056}_{-0.056}$ | $0.0153^{+0.0005}_{-0.0005}$ | $18.71^{+1.90}_{-1.90}$ [†2] | $1.66^{+0.23}_{-0.23}$ | BT15 | B18 |
| LHS 1140 | 15.0 | 8.821 ± 0.024 | M4.5V | 2988 ± 67 | $0.191^{+0.012}_{-0.011}$ | H96, D17, LB20 | c | $3.777929^{+0.000030}_{-0.000030}$ | $1.169^{+0.037}_{-0.038}$ | $0.02734^{+0.00054}_{-0.00054}$ | $5.85^{+0.27}_{-0.25}$ | $1.76^{+0.17}_{-0.17}$ | M19 | LB20 |
| Kepler-78 | 124.0 | 9.586 ± 0.015 | G | 5121 ± 44 | $0.779^{+0.032}_{-0.046}$ | SO13, H13, D19 | b | $0.35500744^{+0.00000006}_{-0.00000006}$ | $1.228^{+0.018}_{-0.019}$ | ⋯ | ⋯ | $1.77^{+0.24}_{-0.25}$ | SO13 | SO13, D19 |
| GJ 357 | 9.4 | 6.475 ± 0.017 | M2.5V | 3505 ± 51 | 0.342 ± 0.011 | H96, S19 | b | $3.93072^{+0.00008}_{-0.00006}$ | $1.217^{+0.084}_{-0.083}$ | $0.035^{+0.002}_{-0.002}$ | $12.6^{+1.1}_{-0.8}$ | $1.84^{+0.31}_{-0.31}$ | L19 | L19 |
| GJ 3473 | 27.4 | 8.829 ± 0.024 | M4.0V | 3347 ± 54 | 0.360 ± 0.016 | H96, K20 | b | $1.1980035^{+0.0000018}_{-0.0000019}$ | $1.264^{+0.050}_{-0.049}$ | $0.01589^{+0.00062}_{-0.00062}$ | $59.4^{+5.9}_{-4.5}$ | $1.86^{+0.30}_{-0.30}$ | K20 | K20 |
| L 98-59 | 10.6 | 7.101 ± 0.018 | M3V±1 | 3415 ± 135 | 0.273 ± 0.030 | K19, D21 | c | $3.6906777^{+0.0000016}_{-0.0000026}$ | $1.385^{+0.095}_{-0.075}$ | $0.0304^{+0.0011}_{-0.0012}$ | $12.8^{+2.6}_{-2.1}$ | $2.22^{+0.26}_{-0.25}$ | K19 | D21 |
| | | | | | | | d | $7.4507245^{+0.0000081}_{-0.0000046}$ | $1.521^{+0.119}_{-0.098}$ | $0.0486^{+0.0018}_{-0.0019}$ | $5.01^{+1.02}_{-0.83}$ | $1.94^{+0.28}_{-0.28}$ | | |
| LHS 1478 | 18.2 | 8.767 ± 0.022 | M3V | 3381 ± 54 | 0.236 ± 0.012 | P13, S21 | b | $1.9495378^{+0.0000040}_{-0.0000041}$ | $1.242^{+0.051}_{-0.049}$ | $0.01848^{+0.00061}_{-0.00063}$ | $20.9^{+1.4}_{-1.4}$ [†3] | $2.33^{+0.20}_{-0.20}$ | S21 | S21 |
| LTT 3780 | 22.0 | 8.204 ± 0.021 | M3.5V | 3360 ± 51 | 0.401 ± 0.012 | R03, C20, N20 | b | $0.768377^{+0.0000014}_{-0.0000014}$ | $1.35^{+0.06}_{-0.06}$ | $0.01203^{+0.00054}_{-0.00053}$ | $116^{+11}_{-10}$ | $2.34^{+0.24}_{-0.23}$ | C20, N20 | N20 |
| K2-229 | 102.6 | 9.050 ± 0.023 | G9 | $5315^{+35}_{-31}$ | $0.87^{+0.01}_{-0.01}$ | Lu18, Li18 | b | $0.58426^{+0.00002}_{-0.00002}$ | $1.197^{+0.045}_{-0.048}$ | $0.0131^{+0.00005}_{-0.00004}$ | $2615^{+98}_{-98}$ [†4] | $2.49^{+0.42}_{-0.43}$ | M18 | Li18, D19 |

**Note.** 1: Bailer-Jones et al. (2018); *: 2MASS (Cutri et al. 2003; Skrutskie et al. 2006); †1: Derived using VE21 $L_\mathrm{bol}$ and $a$; †2: Derived using BT15 $L_\mathrm{bol}$ and B18 $a$; †3: Derived using S21 $L_\mathrm{bol}$ and $a$; †4: Derived using Li18 $T_\mathrm{eff}$, $R_*$, and $a$. Discovery and Property References: H96: Hawley et al. (1996), G00: Gizis et al. (2000), R03: Reid et al. (2003), C06: Costa et al. (2006), H13: Howard et al. (2013), P13: Pecaut & Mamajek (2013), SO13: Sanchis-Ojeda et al. (2013), BT15: Berta-Thompson et al. (2015), D15: Davison et al. (2015), D17: Dittmann et al. (2017), G17: Gillon et al. (2017), B18: Bonfils et al. (2018), Li18: Livingston et al. (2018), Lu18: Luo et al. (2018), M18: Mayo et al. (2018), C19: Cloutier et al. (2019a), D19: Dai et al. (2019), G19: Günther et al. (2019), K19: Kostov et al. (2019a), L19: Luque et al. (2019), M19: Ment et al. (2019), S19: Schweitzer et al. (2019), C20: Cloutier et al. (2020a), K20: Kemmer et al. (2020), LB20: Lillo-Box et al. (2020), N20: Nowak et al. (2020), A21: Agol et al. (2021), D21: Demangeon et al. (2021), S21: Soto et al. (2021), VE21: Van Eylen et al. (2021).

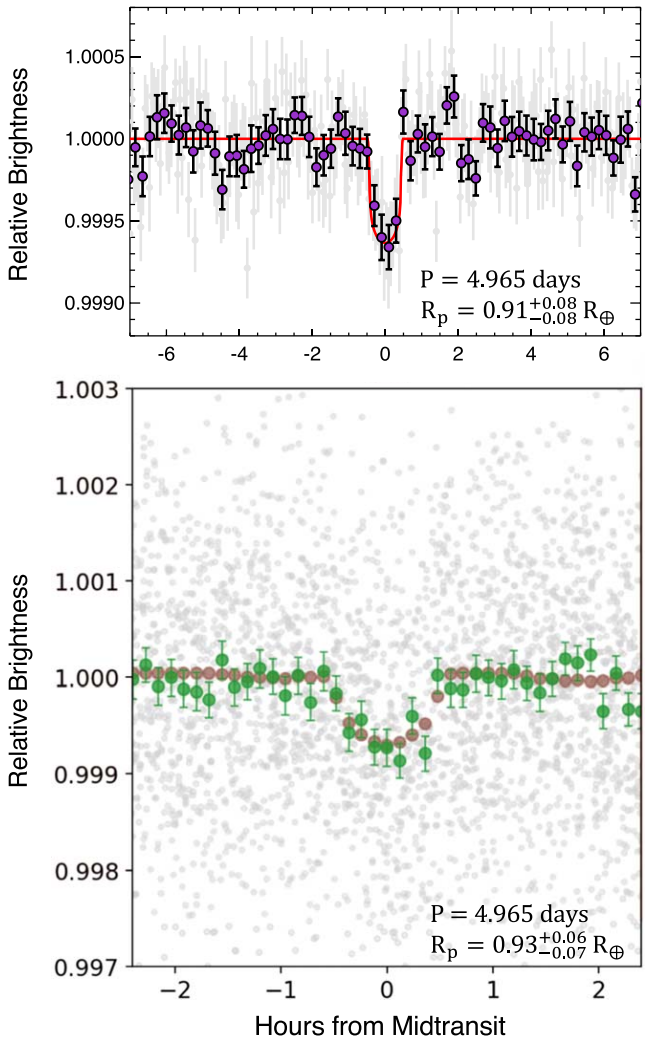

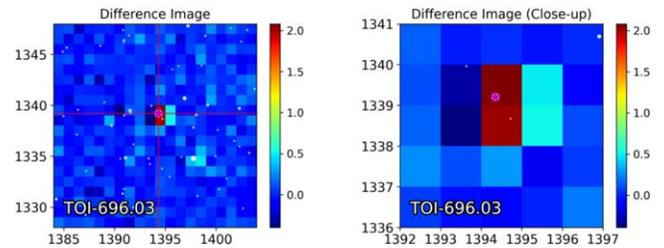

**Figure 19.** Left: the difference between out-of-transit and in-transit images of LHS 1678 star during transits of TOI-696.03 reveal that the star is the location of the candidate transits and rule out other nearby stars as the transit host (small dots). Right: a zoom in on LHS 1678 in the difference image.

**Figure 18.** Phase-folded light curves of TOI-696.03 identified in a search of the full two-cycle, four-sector TESS data set using two independent analyses. Top (Analysis 1): the gray and purple symbols are averages of TESS 2 minute observations in intervals of 3.6 and 12 minutes in phase, respectively. The best-fit model is plotted in red. Bottom (Analysis 2): gray and green symbols represent 2 minute cadence and 7.5 minute binned measurements, respectively. The brown symbols are drawn from the best-fit transit model from this analysis. Both analyses consistently yield a small planet candidate on a 4.965 day period. If confirmed, TOI-696.03 would be sub-Earth-sized and near the 4:3 mean-motion resonance with LHS 1678 c.

with the other bodies in the system. This fits into the picture that most USPs have much larger orbital separation compared to other pairs of planets in the same system (Winn et al. 2018). It is plausible that LHS 1678 b migrated in a resonant chain with the other two planets, and then continued inward toward the star via tidal orbital decay (i.e., Lee & Chiang 2017).

Planets in near-resonant configurations can exhibit TTVs that can reveal the architectures and masses of planets in the system. To predict the amplitude of any putative TTVs in the LHS 1678 system, we estimated the masses of both LHS 1678 c and TOI-696.03 using the forecaster procedure from Chen & Kipping (2017) and found values of $0.940^{+0.606}_{-0.344}$ $M_\oplus$ and $0.72^{+0.57}_{-0.31}$ $M_\oplus$, respectively. Using the relationships of Lithwick et al. (2012), we estimate that both planets should exhibit TTVs with a super-period of about 155 days and amplitudes $\gtrsim 2$ minutes (possibly significantly larger if the planets have any free eccentricity). Ongoing ground-based follow-up of LHS 1678 c and the planned follow-up and validation of TOI-696.03 may reveal TTVs in this system.

### 7.2. Prospects for Future Exoplanet Characterization

Here we explore possibilities for future characterization of the two validated planets in this system, LHS 1678 b and c.

*Masses.* We used HARPS RV monitoring to place stringent upper limits on the masses of LHS 1678 b and c, but lacked the necessary RV precision and observing baseline to significantly measure the masses. Given current knowledge of the mass–radius relation for small planets ($<1.5$ $R_\oplus$), it is likely LHS 1678 b and c are predominantly rocky in nature (Rogers 2015). Figure 17 places these planets in context with other small, low-mass planets. Assuming their expected densities, they join a very small sample of planets comparable in size and mass to the Earth and are compelling targets for further mass constraints to infer their bulk compositions. Using the forecaster procedure from Chen & Kipping (2017), we estimate masses of $0.276^{+0.142}_{-0.101}$ $M_\oplus$ and $0.940^{+0.606}_{-0.344}$ $M_\oplus$ for planets b and c, respectively (matching the estimate using the relation in Kempton et al. 2018). These masses are broadly consistent with our mass upper limits (Table 7, Figure 17). They are also below the collisional stripping mass limits (0.419 $M_\oplus$ and 1.51 $M_\oplus$ for b and c, respectively) as described by Marcus et al. (2010) and as plotted in Figure 17 using the table from Zeng et al. (2019).[60] With the forecaster masses, assuming circular orbits around LHS 1678, and taking the remaining parameters from Tables 2 and 7, we derive radial velocity signals of approximately 0.4 and 0.8 m s$^{-1}$. Given the predicted amplitude of these signals and LHS 1678's southern decl., ESPRESSO on the VLT is a current instrument with demonstrated performance capable of better refining the planet masses (Pepe et al. 2010; Suárez Mascareño et al. 2020). Definitive mass measurements of these planets will be critical for future bulk composition studies and atmospheric characterization.

*Atmospheric characterization.* Kreidberg et al. (2019) demonstrated the use of thermal phase curves to investigate the atmospheres of small rocky planets when they inferred from Spitzer observations that the USP LHS 3844 b, which orbits its host star in only 11 hr (Vanderspek et al. 2019), likely has no atmosphere. With an orbit of 0.86 days, LHS 1678 b falls into the USP category. Because it is likely rocky, nearby, and orbiting an even brighter M dwarf ($V_J = 12.5$, $K_s = 8.3$ for LHS 1678 versus $V_J = 15.3$, $K_s = 9.1$ for LHS 3844; Cutri et al. 2003; Skrutskie et al. 2006; Winters et al. 2015; Vanderspek et al. 2019), it provides another opportunity to investigate the

---
[60] https://www.cfa.harvard.edu/~lzeng/planetmodels.html#mrtables

Table 9
Preliminary TOI-696.03 Parameters

| Parameter | Median | 1σ | Median | +1σ | −1σ |
|---|---|---|---|---|---|
| | Analysis 1 | | Analysis 2 | | |
| *Sampled Parameters* | | | | | |
| Limb darkening TESS $u_1$ | 0.20[a] | 0.12 | 1.26 | 0.15 | 0.17 |
| Limb darkening TESS $u_2$ | 0.45[a] | 0.14 | −0.45 | 0.17 | 0.13 |
| Time of reference transit ($T_0$) [BJD-2457000] | 1806.8143 | 0.0015 | 1414.5629 | 0.0015 | 0.0012 |
| Period [days] | 4.965222 | 0.000023 | 4.965207 | 0.000011 | 0.000012 |
| $R_p/R_*$ | 0.0255 | 0.0022 | 0.0261 | 0.0015 | 0.0018 |
| Impact parameter | 0.807 | 0.060 | 0.716 | 0.035 | 0.055 |
| *Derived Parameters* | | | | | |
| $a$ [au] | 0.03995 | 0.00054 | 0.03996 | 0.00026 | 0.00026 |
| $a/R_*$ | 26.06 | 0.90 | 26.09 | 0.39 | 0.38 |
| Radius [$R_\oplus$] | 0.915 | 0.085 | 0.931 | 0.064 | 0.067 |
| Inclination [deg] | 88.22[b] | 0.16 | 89.6 | 1.9 | 1.1 |
| Duration [hours] | 0.924 | 0.094 | 1.068 | 0.067 | 0.049 |
| Insolation [$S_\oplus$] | 9.09 | 0.31 | 9.08 | 0.36 | 0.35 |

**Notes.**
[a] Note that these values are dominated by the imposed priors in Analysis 1.
[b] Note that the inclination in Analysis 1 is forced to a value below 90 deg.

atmosphere of a small planet in an extreme orbital architecture and radiation environment. Such planets provide key constraints on the evolution of small-planet atmospheres under the influence of low-mass star space weather. Although LHS 1678 is now magnetically quiet, it was certainly active as a young star, and may have eroded or fully stripped away one or both of its planets' atmospheres.

With an insolation flux of 93.2 $S_\oplus$ and an estimated equilibrium temperature in excess of 700 K, assuming albedos ranging from 0.0 to 0.7, LHS 1678 b may be sufficiently hot for thermal phase curve measurements using the JWST. We use the emission spectroscopy metric (ESM; Kempton et al. 2018) to investigate the planet's potential for such observations. The planet has an ESM of 3.9, indicating that three full phase curve observations with the JWST MIRI LRS mode would be required for a significant constraint on its atmosphere. However, in this case, the ESM value undersells the potential of LHS 1678 b observations using the JWST. We note that under broader assumptions of the planet's properties, the thermal emission may be significantly higher than predicted by the ESM. For example, in the case that the planet's dayside is bare rock and a tenuous or nonexistent atmosphere provides very inefficient heat redistribution (similar to LHS 3844 b), the dayside temperature could be significantly higher (>1000 K) and lead to thermal emission detectable by the JWST MIRI in a single phase curve observation. Thus, LHS 1678 b is a compelling target for future JWST observations and may become a benchmark in the study of small-planet atmospheres under the influence of extreme radiation. We also calculated an ESM of 2.0 for LHS 1678 c. This cooler, longer-period planet is a less compelling target for thermal emission measurements than its shorter-period, hotter sibling.

We also investigated both planets' prospects for transmission spectroscopy observations using the Kempton et al. (2018) transmission spectroscopy metric (TSM). We do this using two sets of possible masses. Using our 2-sigma mass upper limits ($M_b = 0.35\,M_\oplus$, $M_c = 1.4\,M_\oplus$), we derive TSM lower limits of 22.9 and 9.9 for planets b and c; using masses calculated as described in Kempton et al. (2018; $M_b = 0.27\,M_\oplus$, $M_c = 0.91\,M_\oplus$), we derive TSMs of 30.2 and 15.2. Kempton et al. (2018) assert that planets with TSM > 10 are compelling targets for atmospheric characterization via transmission spectroscopy measurements; this TSM threshold corresponds to a planet c mass of ≲1.4 $M_\oplus$, our 2-sigma mass upper limit.

We point out that with an insolation flux of 93.2 $S_\oplus$, planet b exists in an extreme radiation environment and may have no atmosphere to measure in transmission. Alternatively, it could have retained a high-mean molecular weight atmosphere, which would have a small-scale height and produce a small transmission signal. Given its significantly lower insolation flux, planet c may be a more compelling transmission spectroscopy target. In trying to understand the runaway greenhouse effect, it is important to assemble a suite of planets at different stages in the process. With an insolation flux of 13.5 $S_\oplus$ and radius of 0.982 $R_\oplus$, planet c falls in the middle of the VZ range defined by Kane et al. (2014). The outer limit of the VZ is defined by the distance from the star at which one expects sufficient erosion of the atmosphere to counteract the runaway greenhouse effect (∼1 $S_\oplus$ around a 3500 K star). The inner limit of (∼25 $S_\oplus$) is set by a predicted complete evaporation of oceans. Planet c is not only in the VZ, but is the same radius as Venus within uncertainties and likely rocky, given its radius and our mass upper limits. These qualities make it an especially compelling target for inclusion in runaway greenhouse effect studies, and its TSM lower limit of 9.9 speaks to its feasibility.

### 7.3. The Nature of the Wide-orbit Companion

RECONS long-baseline astrometry provides compelling evidence for a wide-orbit, low-mass companion in the LHS 1678 system (see Section 3.5). However, the companion has not yet been directly detected. In Appendix C we discuss in greater depth our constraints on the companion. In combination, the available observations provide constraints on its mass and luminosity that indicate it is likely a brown dwarf with a projected separation of ≲5 au, while not disallowing the possibility it is a Jovian planet. Because the astrometric perturbation is so far seen exclusively in the decl. axis, the orbit is highly inclined, and the semimajor axis of the system may be larger than the projected separation upper limit placed by the imaging observations (Section C.2).

One additional intriguing possibility is worth noting: the high inclination of the companion's orbit also suggests the possibility that the transiting planets and companion are coplanar. With this in mind, additional astrometric data can constrain the inclination of the (so far) unseen companion's orbit, perhaps revealing a rare system in which companions of very different masses orbit in the same plane. This would be the second case in the solar neighborhood, following that found for the nearby (6.9 pc) M dwarf triple LTT 1445 (Winters et al. 2019), in which the A component harbors at least one short-period transiting planet and the BC pair appear to orbit one another in the same plane. LHS 1678 may join a growing ensemble of orbitally aligned systems and be key to understanding their frequency and the mechanisms that drive their formation and evolution (Christian et al. 2022).

### 7.4. System Stability

Studying the dynamics of planetary systems can validate the viability of the measured Keplerian orbital solution and reveal potential locations for additional planets in the system (Lissauer et al. 2011; Kane & Raymond 2014; Li et al. 2014; Kane & Blunt 2019). To test the dynamical aspects of the orbital architecture of the LHS 1678 system, we performed $N$-body integrations using the Mercury Integrator Package (Chambers 1999) and the system parameters in Tables 2 and 7. The methodological approach taken was similar to that described by Kane (2015, 2019). We adopted a time step in integrations of 0.04 days, necessitated by the very short orbital period of the inner planet. A single simulation was allowed to run for $10^7$ simulation years and demonstrated that the system as described is exceptionally stable, with minimal perturbative interactions between the planets. We note that because the astrometric companion would not affect the planetary system unless highly eccentric (Kane 2019), we excluded it in this simulation. The combination of the relatively small predicted masses of the planets and their proximity to the host star results in similar small Hill radii: $1.1 \times 10^{-4}$ au and $4.2 \times 10^{-4}$ au for planets b and c, respectively. These small Hill radii imply that a significant amount of viable orbital space lies between the planets where additional, perhaps nontransiting, planets could reside. To test for this, we conducted several hundred $N$-body integrations that randomly inserted an Earth-mass planet at locations between the two known planets and ran each integration for $10^5$ simulation years. These simulations showed that an additional terrestrial planet is dynamically feasible in the semimajor axis range 0.014–0.029 au, as is the presence of candidate planet TOI-696.03 at 0.04 au (Section 7.1). Statistical studies of compact planetary systems show that they are generally dynamically filled (Fang & Margot 2013). It is therefore possible that there are additional planets within the system that are interior to the outer detected planet.

Note that we did not include TOI-696.03 in our dynamical simulations due to its candidate status and lack of mass constraints, other than to infer stability at its tentatively derived location. Given its estimated radius of $\sim 0.9\, R_\oplus$ and $\sim 0.04$ au, and assuming a mass of $0.9\, M_\oplus$, TOI-696.03 would have a similarly small Hill radius of $5.5 \times 10^{-4}$ au. Such a small mass and Hill radius is unlikely to have significant impact on the stability derived between planets b and c, depending on the eccentricity of the orbit for TOI-696.03 and locations of possible resonances (Hadden & Lithwick 2018; Vinson & Chiang 2018; Hadden 2019; Kane et al. 2021). Further validation of TOI-696.03 and measurements that constrain its mass and orbital properties will allow a more thorough dynamical investigation into the architecture of the system.

### 7.5. Potential Impacts of Association with a Gap in Lower Main Sequence

LHS 1678's HR-diagram position in or at the lower edge of an observed gap in the lower main sequence is nearly unique among TESS exoplanet host stars. The gap is associated with a core $^3$He instability that leads to damped, periodic pulsations in stellar radius and luminosity on the order of a few to $\sim 10\%$ (van Saders & Pinsonneault 2012; MacDonald & Gizis 2018; Baraffe & Chabrier 2018). Depending on the mass of the star, these pulsations occur on timescales of $\sim 10^8$ to $10^9$ yr. Changes in stellar luminosity of a few to $\sim 10\%$ may have an impact on the evolution of a close-in planet, particularly its atmosphere and surface temperature. The effect of host star luminosity changes on exoplanet climates is an open question. It is conceivable that such changes in incident radiation could push planets in appropriate parts of parameter space toward different climate states, like runaway greenhouse and surface desiccation, or in the direction of a habitable state. This is highly dependent on the planet parameters, including period and atmospheric content. M dwarf planets often present an additional challenge to understanding these effects; they are often tidally locked, which adds further complexity to their atmospheres (e.g., Yang et al. 2013; Kopparapu et al. 2016). Although the implications of such instellation changes on exoplanets need further study, they have been explored in the context of Venus and Earth and solar evolution. Modeling has shown that a ≳10% increase in solar luminosity could trigger an abrupt shift in the Earth's climate in the direction of a moist greenhouse and significant surface warming (Wolf & Toon 2015). These implications, and the unique stellar evolution history of the LHS 1678 system, provide the impetus to model and explore the impact of periodic stellar luminosity changes on the properties of planets orbiting HR-diagram-gap M dwarfs. This may be particularly important for LHS 1678 c, given its location in the VZ.

### 8. Conclusions

The LHS 1678 system stands out among other high-profile TESS discoveries. The star is less than 20 pc from the Sun and bright in both visible and IR wavelengths. Both validated planets are smaller than Earth, very likely rocky, and each is interesting in its own right: LHS 1678 b is an USP planet with prospects for thermal emission measurements, and LHS 1678 c is a Venus analog. The additional planet candidate in the system, TOI-696.03, is near the 4:3 mean-motion resonance with LHS 1678 c and may lead to measurable TTVs in the system with further follow-up. LHS 1678 is associated with the lower main sequence gap revealed by Gaia (Jao et al. 2018). This gap is tied to low-amplitude, long-period luminosity oscillations stemming from a core $^3$He burning instability. The effects of these oscillations on exoplanet formation and evolution are unknown. LHS 1678 also has a very-low-mass, astrometrically detected substellar companion with an orbit on the order of decades. LHS 1678 complements other low-mass multiple systems where one component is known to host planets, like Kepler-296 (Barclay et al. 2015), K2-288 (Feinstein et al. 2019), and LTT 1445 (Winters et al. 2019). A complete census of each exoplanet system is key to understanding its formation and evolution. The nondetection of this companion by photometric and spectroscopic measurements highlights the difficulty in probing the full phase space of companions at the star/brown dwarf boundary and the value of long-term observing programs. The full characterization presented here and the combination of the traits revealed make the LHS 1678 system a compelling target for numerous future follow-up studies.

We thank Daniel Foreman-Mackey for valuable suggestions on sampling parameterization.

This publication makes use of data products from the Two Micron All Sky Survey, which is a joint project of the University of Massachusetts and the Infrared Processing and Analysis Center/California Institute of Technology, funded by the National Aeronautics and Space Administration and the National Science Foundation.

This publication makes use of data products from the Wide-field Infrared Survey Explorer, which is a joint project of the University of California, Los Angeles, and the Jet Propulsion Laboratory/California Institute of Technology, funded by the National Aeronautics and Space Administration.

This work has made use of data from the European Space Agency (ESA) mission Gaia (https://www.cosmos.esa.int/gaia), processed by the Gaia Data Processing and Analysis Consortium (DPAC, https://www.cosmos.esa.int/web/gaia/dpac/consortium). Funding for the DPAC has been provided by national institutions, in particular the institutions participating in the Gaia Multilateral Agreement.

This paper includes data collected with the TESS mission (Ricker et al. 2015), obtained from the MAST data archive at the Space Telescope Science Institute (STScI). Funding for the TESS mission is provided by NASAs Science Mission Directorate. STScI is operated by the Association of Universities for Research in Astronomy, Inc., under NASA contract NAS 526555.

We acknowledge the use of TESS High Level Science Products (HLSP) produced by the Quick-Look Pipeline (QLP) at the TESS Science Office at MIT, which are publicly available from the Mikulski Archive for Space Telescopes (MAST). Funding for the TESS mission is provided by NASA's Science Mission directorate.

We acknowledge the use of public TESS data from pipelines at the TESS Science Office and at the TESS Science Processing Operations Center. Resources supporting this work were provided by the NASA High-End Computing (HEC) Program through the NASA Advanced Supercomputing (NAS) Division at Ames Research Center for the production of the SPOC data products.

This research has made use of the Exoplanet Follow-up Observation Program website, which is operated by the California Institute of Technology, under contract with the National Aeronautics and Space Administration under the Exoplanet Exploration Program.

This work makes use of observations from the LCOGT network. Part of the LCOGT telescope time was granted by NOIRLab through the Mid-Scale Innovations Program (MSIP). MSIP is funded by NSF.

This research has made use of "Aladin sky atlas", developed at CDS, Strasbourg Observatory, France (Bonnarel et al. 2000), the SIMBAD database, operated at CDS, Strasbourg, France (Wenger et al. 2000), and the VizieR catalog access tool, CDS, Strasbourg, France (Ochsenbein et al. 2000).

This research made use of the TOPCAT database visualization and management software, available at http://www.star.bris.ac.uk/~mbt/topcat/ (Taylor 2005).

This research made use of Lightkurve, a Python package for Kepler and TESS data analysis (Lightkurve Collaboration et al. 2018).

This work made use of tpfplotter by J. Lillo-Box (publicly available in www.github.com/jlillo/tpfplotter), which also made use of the Python packages astropy, Lightkurve, matplotlib and numpy.

This research made use of exoplanet (Foreman-Mackey et al. 2021) and its dependencies (Agol et al. 2020; Kumar et al. 2019; Astropy Collaboration et al. 2013, 2018; Kipping 2013; Luger et al. 2019; Salvatier et al. 2016; Theano Development Team 2016).

The MEarth Team gratefully acknowledges funding from the David and Lucile Packard Fellowship for Science and Engineering (awarded to D.C.). This material is based upon work supported by the National Science Foundation under grant Nos. AST-0807690, AST-1109468, AST-1004488 (Alan T. Waterman Award), and AST-1616624, and upon work supported by the National Aeronautics and Space Administration under grant No. 80NSSC18K0476 issued through the XRP Program. This work is made possible by a grant from the John Templeton Foundation. The opinions expressed in this publication are those of the authors and do not necessarily reflect the views of the John Templeton Foundation.

N. A.-D. acknowledges the support of FONDECYT project 3180063.

T.D. acknowledges support from MIT's Kavli Institute as a Kavli postdoctoral fellow.

K.H. acknowledges support from STFC grant No. ST/R000824/1.

E.A.G. thanks the LSSTC Data Science Fellowship Program, which is funded by LSSTC, NSF Cybertraining Grant #1829740, the Brinson Foundation, and the Moore Foundation; her participation in the program has benefited this work. The material is based upon work supported by NASA under award number 80GSFC21M0002.

This work was supported by the lead author's appointment to the NASA Postdoctoral Program at the Goddard Space Flight Center, administered by Universities Space Research Association under contract with NASA.

*Facilities:* TESS, CTIO:0.9 m, CTIO:1.5 m, WISE, Small and Moderate Aperture Research Telescope System (SMARTS), LCOGT, PEST, MEarth, VLT NaCo, SOAR HRCam and SAM, Gaia, HARPS, 2MASS, MEarth-South.

*Software:* Python, IDL, NumPy (Harris et al. 2020), Matplotlib (Hunter 2007), Astropy (Astropy Collaboration et al. 2013, 2018), Lightkurve (Lightkurve Collaboration et al. 2018), DAVE (Kostov et al. 2019b), QATS, (Carter & Agol 2013; Kruse et al. 2019), TLS (Hippke & Heller 2019a, 2019b), AstroImageJ (Collins et al. 2017), vespa (Morton 2015), CDS aladin (Bonnarel et al. 2000), SIMBAD (Wenger et al. 2000) and VizieR (Ochsenbein et al.

2000), tpfplotter (Aller et al. 2020), exoplanet (Foreman-Mackey et al. 2021; Agol et al. 2020; Kumar et al. 2019; Astropy Collaboration et al. 2013, 2018; Kipping 2013; Luger et al. 2019; Salvatier et al. 2016; Theano Development Team 2016), forecaster (Chen & Kipping 2017), TESS-cut (Brasseur et al. 2019), Quick-Look Pipeline (Huang et al. 2020a, 2020b), TESS-plots (https://github.com/mkunimoto/TESS-plots).

# Appendix A
## Stellar Properties—Multiple Techniques

We use multiple methods to robustly determine the stellar properties of LHS 1678, with the adopted methods described in Section 3.2. Here we describe the other methods in greater depth, with results in Table 10.

For one alternate method, we employed the $\mathcal{M}_{Ks}$-based relations of Mann et al. (2015, 2019) to estimate the mass, radius, and luminosity of LHS 1678. We included in these calculations the effect of stellar metallicity using a value of [Fe/H] = −0.5 for LHS 1678. We then combined the luminosity and radius using the Stefan–Boltzmann law to estimate the effective temperature. The estimated mass uncertainty follows from the probabilistic method described in Mann et al. (2019). We estimated uncertainties on the radius, luminosity, and temperature via Monte Carlo methods assuming Gaussian measurement errors and added the systematic scatter in the Mann et al. (2015) relations in quadrature. We found stellar parameters consistent with those adopted for the analyses presented in the text and estimated using the methods of M. L. Silverstein et al. (2022, in preparation). The Mann et al. parameters are also consistent with the estimated spectral type of LHS 1678.

We also performed an analysis of the broadband SED of LHS 1678 together with the Gaia DR2 parallaxes (adjusted by +0.08 mas to account for the systematic offset reported by Stassun & Torres 2018), in order to determine an empirical measurement of the stellar radius, following the procedures described in Stassun & Torres (2016) and Stassun et al. (2017, 2018a). We pulled the $U$ magnitude from Mermilliod (2006), the $JHK_s$ magnitudes from 2MASS, the W1–W4 magnitudes from the WISE All-Sky Data Release (Wright et al. 2010; Cutri et al. 2012), the $G\,BP\,RP$ magnitudes from Gaia DR2, and the NUV magnitude from the Galaxy Evolution Explorer (GALEX; Martin et al. 2005; Bianchi et al. 2011). Together, the available photometry spans the full stellar SED over the wavelength range 0.2–22 $\mu$m (see Figure 20). We exclude the GALEX point in our analysis because there appears to be excess flux, likely due to chromospheric activity commonly found in M dwarfs. All values are in Table 1, except for WISE All-Sky Data Release W1 = 8.117 ± 0.024, W2 = 7.972 ± 0.021, W3 = 7.877 ± 0.017, W4 = 7.823 ± 0.117. We switch from the WISE AllWISE to All-Sky release here because we make use of W4 band, which has a higher S/N in the All-Sky release.

We performed a fit using NextGen stellar atmosphere models (Hauschildt et al. 1999), with the free parameters being the effective temperature ($T_{\rm eff}$) and metallicity ([Fe/H]). The broadband SED is largely insensitive to the choice of log $g$, therefore we adopted a value of 5.0 as expected for an M dwarf. We also fixed the extinction $A_V \equiv 0$ based on the star's proximity. The resulting fit (Figure 20) corresponds to $T_{\rm eff} = 3550 \pm 100$ K and [Fe/H] = −0.5, and has a reduced $\chi^2$ of 3.3. Integrating the (unreddened) model SED gives the bolometric flux at Earth, $F_{\rm bol} = 1.158 \pm 0.045 \times 10^{-9}$ erg s$^{-1}$ cm$^{-2}$. Taking the $F_{\rm bol}$ and $T_{\rm eff}$ together with the Gaia DR2 parallax gives the stellar radius, $R_\star = 0.316 \pm 0.018 R_\odot$. These parameter values are also consistent with those determined using the methods of M. L. Silverstein et al. (2022, in preparation).

Table 10
Host Star Derived Properties

| Property | Value | Error | Ref. |
| --- | --- | --- | --- |
| Names | LHS 1678, TIC 77156829, TOI-696, L 375-2, LTT 2022, NLTT 13515 | | |
| Mass ($M_\odot$) | **0.345** | **0.014** | Avg. of Benedict et al. (2016) $\mathcal{M}_V$ and $\mathcal{M}_K$ Relations |
| | 0.348 | 0.021 | Benedict et al. (2016) $\mathcal{M}_K$ Relation |
| | 0.341 | 0.020 | Benedict et al. (2016) $\mathcal{M}_V$ Relation |
| | 0.322 | 0.009 | Mann et al. (2019) Relation with [Fe/H] = −0.5 Term |
| | 0.345 | 0.028 | Mann et al. (2019) Relation with No [Fe/H] Term |
| Effective temperature ($K$) | **3490** | **50** | This work: Section 3.2 (M. L. Silverstein et al. 2022, in preparation) |
| | 3420 | 160 | Mann et al. (2015) Relation with [Fe/H] = −0.5 Term |
| | 3550 | 100 | NextGen SED Method[a] |
| Bolometric flux (erg cm$^{-2}$ s$^{-1}$) | **1.171 × 10$^{-9}$** | 0.022 × 10$^{-9}$ | This work: Section 3.2 (M. L. Silverstein et al. 2022, in preparation) |
| | 1.158 × 10$^{-9}$ | 0.045 × 10$^{-9}$ | NextGen SED Method[a] |
| Luminosity ($L_\odot$) | **0.0145** | **0.0003** | This work: Section 3.2 (M. L. Silverstein et al. 2022, in preparation) |
| | 0.0137 | 0.0008 | Mann et al. (2015) Relation with [Fe/H] = −0.5 Term |
| Radius ($R_\odot$) | **0.329** | **0.010** | This work: Section 3.2 (M. L. Silverstein et al. 2022, in preparation) |
| | 0.333 | 0.030 | Mann et al. (2015) Relation with [Fe/H] = −0.5 Term |
| | 0.316 | 0.018 | NextGen SED Method[a] |

**Note.** Adopted values in bold.
[a] As described in Stassun & Torres (2016); Stassun et al. (2017, 2018a).

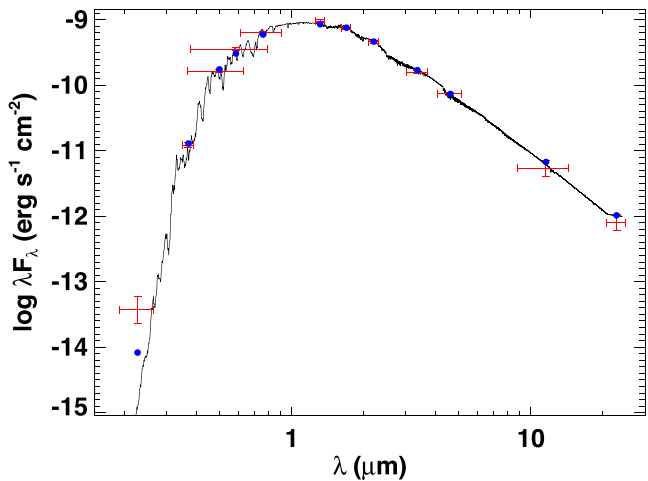

**Figure 20.** SED of LHS 1678 with best-fit stellar model used to determine an alternate set of stellar parameters. Red symbols represent the observed photometric measurements, where the horizontal bars represent the effective width of the passband. Blue symbols are the model fluxes from the best-fit NextGen atmosphere model (black).

We reiterate that the three methods used in this paper to derive stellar effective temperature, luminosity, and radius fall into better agreement when all three assume subsolar metallicity, which is consistent with other properties of LHS 1678 (e.g., low HR-diagram position and large galactic velocities). We note that the stellar parameter analyses presented in Rains et al. (2021) also provide consistent results and a similar subsolar metallicity assessment.

## Appendix B
## TESS M Dwarf Exoplanet Systems

Here in Table 11, we list the names of the M dwarf exoplanet systems presented in Figure 3 and their discovery references.

**Table 11**
TESS M Dwarf Exoplanet Systems as of 2021 October 20, as in Figure 3

| System | Discovery Reference(s) |
|---|---|
| TOI-1749 | Fukui et al. (2021) |
| TOI-1235 | Cloutier et al. (2020b) |
| TOI-1728 | Kanodia et al. (2020) |
| TOI-1899 | Cañas et al. (2020) |
| TOI-532 | Kanodia et al. (2021) |
| LP 714-47 | Dreizler et al. (2020) |
| LP 961-53 (TOI-776) | Luque et al. (2021) |
| L 168-9 | Astudillo-Defru et al. (2020) |
| AU Mic | Plavchan et al. (2020), Martioli et al. (2021) |
| TOI-1266 | Demory et al. (2020) |
| LHS 1815 | Gan et al. (2020) |
| TOI-1231 | Burt et al. (2021) |
| LHS 1678 | This Work |
| GJ 1252 | Shporer et al. (2020) |
| L 231-32 (TOI-270) | Günther et al. (2019) |
| TOI-1201 | Kossakowski et al. (2021) |
| TOI-700 | Gilbert et al. (2020) |
| GJ 357 | Luque et al. (2019) |
| TOI-1685 | Hirano et al. (2021); Bluhm et al. (2021) |
| TOI-1634 | Cloutier et al. (2021); Hirano et al. (2021) |
| L 98-59 | Kostov et al. (2019a); Demangeon et al. (2021) |
| LHS 1478 (TOI-1640) | Soto et al. (2021) |
| TOI-674 | Murgas et al. (2021) |
| HATS-71 | Bakos et al. (2020) |
| LTT 1445 A | Winters et al. (2019); Winters et al. (2021)[a] |
| TOI-122 | Waalkes et al. (2021) |
| LTT 3780 (TOI-732) | Cloutier et al. (2020a); Nowak et al. (2020) |
| LHS 1972 (GJ 3473, TOI-488) | Kemmer et al. (2020) |
| GJ 486 | Trifonov et al. (2021) |
| TOI-237 | Waalkes et al. (2021) |
| TOI-2406 | Wells et al. (2021) |
| TOI-540 | Ment et al. (2021) |
| LHS 3844 | Vanderspek et al. (2019) |
| LP 791-18 | Crossfield et al. (2019) |

**Note.**
[a] On arXiv, accepted for publication in AJ.

## Appendix C
## Limits on the Properties of the Astrometric Companion

Here we provide the details of our approach to observationally constrain the properties of the astrometrically detected wide-orbit companion described in Sections 3.5 and 7.3. All astrometry used in this discussion comes from the RECONS team, whose data and analysis reveal the companion.

### C.1. Limits Placed by Astrometry

The astrometric perturbation shown in Figure 4 is caused by movement of the photocenter of the unresolved binary orbiting its center of mass. To constrain the relative orbit between the two components, the photocentric orbit can be fit to yield the photocentric semimajor axis, $\alpha_{\text{observed}}$, which in turn can be used to estimate the mass of the unseen companion.

Following the prescription of Van De Kamp (1967), $\alpha_{\text{observed}}$ can be derived using the relative semimajor axis ($a$), along with the masses of the two components ($M_A$, $M_B$), and their fluxes ($F_A$, $F_B$), or apparent magnitudes ($m_A$, $m_B$). More specifically,

$$\alpha_{\text{observed}} = (B - \beta)a, \quad (C1)$$

where $B$ is the fractional mass and $\beta$ is the fractional flux:

$$B = \frac{M_B}{M_A + M_B}$$
$$\beta = \frac{F_B}{F_A + F_B} = \frac{1}{1 + 10^{0.4 \times \Delta m}}. \quad (C2)$$

Here, $\Delta m$ is the magnitude difference between the primary and secondary in the filter used for the astrometry, in this case, $V$. From Kepler's third law, the semimajor axis of the relative orbit, $a$, of a binary with orbital period $P$ is derived from

$$a = P^{2/3}(M_A + M_B)^{1/3}. \quad (C3)$$

By substituting different values of $B$, $\beta$, and $a$ into the relation above (Equation (C1)), we can try to reproduce the observed semimajor axis of the photocentric orbit of LHS 1678 and its companion. The successes/failures constrain the properties we substitute in. We proceed as follows:

1. Simulated primary and secondary masses are allowed to range from 0.600 to 0.002 $M_\odot$. We adopt a step size of 0.010 $M_\odot$ from 0.600 $M_\odot$ to 0.010 $M_\odot$ and 0.002 $M_\odot$ from 0.010 $M_\odot$ down to 0.002 $M_\odot$. This sets the value of $B$.
2. We use the empirical M dwarf mass–luminosity relation (MLR) of Benedict et al. (2016) to convert each primary and secondary mass adopted above to an absolute magnitude $\mathcal{M}_V$, and calculate $\Delta m$. The $\mathcal{M}_V$ values for both components are then merged into a single value for the combined light, and the total apparent magnitude ($V_{total}$) of the two components is then derived using the parallax. For secondary masses less than 0.080 $M_\odot$, which is the limit of the MLR relation and approximately corresponds to the transition between stars and brown dwarfs, we assume the companion contributes no flux at $V$. This sets the value of $\beta$.
3. For a given $M_A$ and $M_B$, we then calculate the relative semimajor axis $a$ using Kepler's third law and an adopted period compatible with the observed photocentric motion. This sets the value of $a$.
4. Using the assigned fractional mass ($B$), fractional flux ($\beta$), and semimajor axis of the relative orbit ($a$), we determine the value of $\alpha_{\text{simulated}}$ for each simulated system.

The results of the simulated photocentric semimajor axis ($\alpha_{\text{simulated}}$) values are shown in Figure 21. Note that because the primary star is assumed to be the more massive component, the top half of each plot is not a physical solution and is shaded black.

We use the available mass and flux constraints as follows to determine the best match between the $\alpha_{\text{simulated}}$ and $\alpha_{\text{observed}}$ (red boxes in Figure 21). (1) Key to the simulations is the mass of the primary, which we estimate using available photometry, the RECONS parallax, and the mass–luminosity relations in Benedict et al. (2016). We estimate $M = 0.38\,M_\odot$ using $\mathcal{M}_V = 11.07$, adopting an uncertainty of $\pm 0.10\,M_\odot$ to allow for the lack of age and metallicity considerations in the Benedict et al. (2016) MLR (all masses below use these relations). (2) We allow the simulated $V_{\text{total}}$ to fall within a range of $12.48 \pm 0.10$, the observed magnitude of the system. (3) Finally, we introduce a third constraint that the difference between $\alpha_{\text{simulated}}$ and $\alpha_{\text{observed}}$ must be 10 mas or less for the match to be deemed successful.

Note that because both the orbital period and $\alpha_{\text{observed}}$ are uncertain, we opt to explore several specific options in Figure 21. We select orbital periods of 42 yr (the value from the fit in Figure 4) and 100 yr, and photocentric semimajor axes of 17 mas (from the fit) and 100 mas. The resulting mass combinations consistent with the astrometry are enclosed in red boxes in Figure 21, where the mass of the primary is shown on the $y$-axis and the mass of the astrometric companion is shown on the $x$-axis. Shading indicates the semimajor axis of the resulting photocentric orbit.

The two left panels of Figure 21 show that for a semimajor axis of the photocentric orbit of 17 mas, the companion is a low-mass brown dwarf if the orbital period is 42 yr, and an even-lower-mass brown dwarf if the period is 100 yr. On the other hand, as shown in the lower-right panel, if the semimajor axis is 100 mas and the orbital period is a century, the companion is a more massive brown dwarf. Note that simulations of a 42 yr orbit and 100 mas semimajor axis shown in the upper-right panel yield a null result: there are no mass combinations consistent with the available astrometric data. *Although the nature of the companion remains uncertain, we find that it is much more likely to be a brown dwarf than a low-mass star, and we cannot yet eliminate the possibility that it is a Jovian planet.*

### C.2. Limits Placed by Imaging

The nature of the companion may also be constrained using photometric and high-resolution imaging data.[61] Using *VRIJHK* photometry from the CTIO/SMARTS 0.9 m and 2MASS, and the relations of Henry et al. (2004), we derive a photometric distance estimate of $27.1 \pm 4.2$ pc. This is much further than the distance of $19.88 \pm 0.01$ pc (Bailer-Jones et al. 2018). Thus, the photometric data appear to eliminate a companion that is similar in mass and luminosity to LHS 1678 because it appears significantly *underluminous* rather than overluminous.

---

[61] Available TESS and ground-based light curves do not provide any additional information about the nature of the companion.

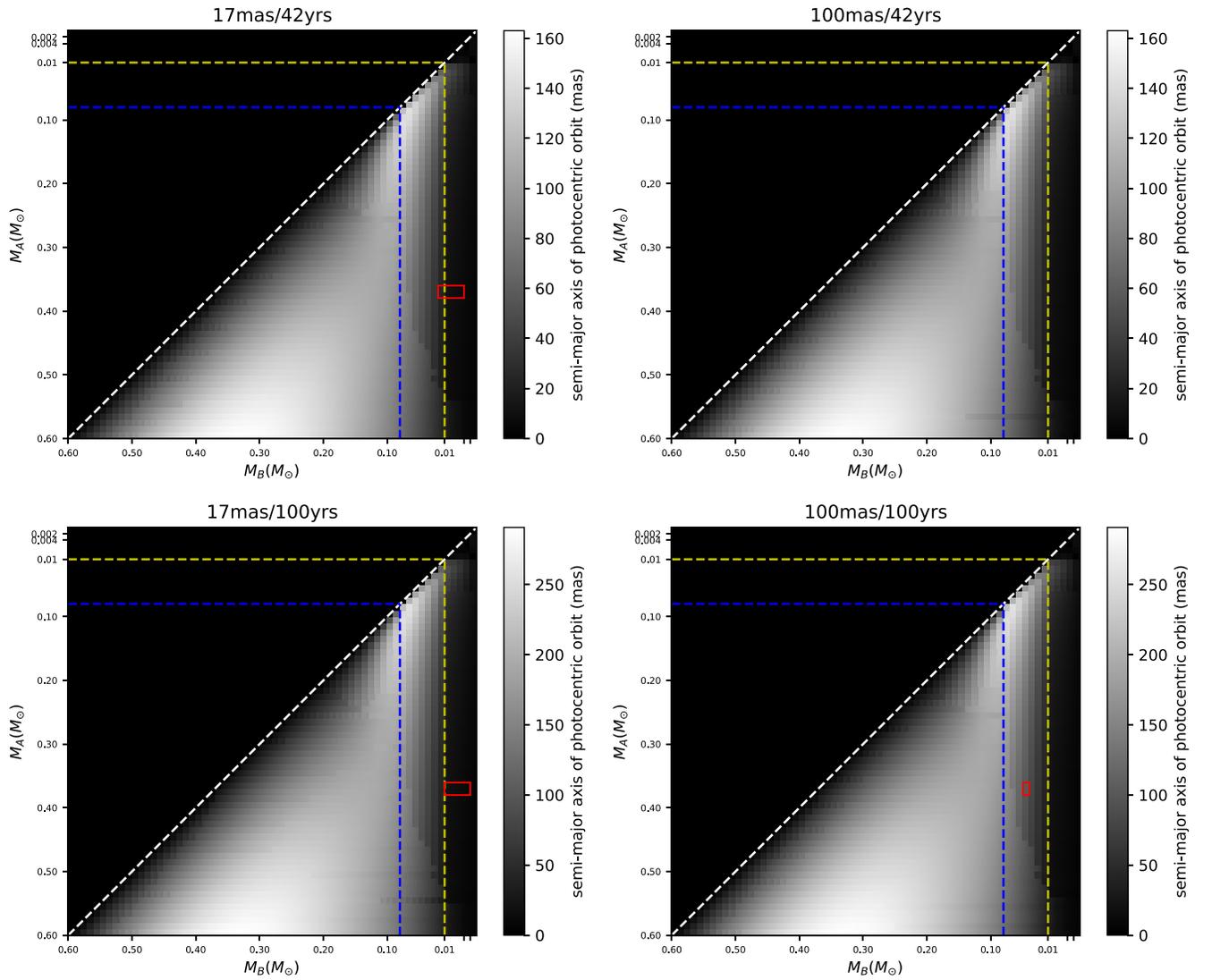

**Figure 21.** Simulated semimajor axes of a photocentric orbit for LHS 1678 based on RECONS astrometry data. Four different scenarios are simulated based on the assumptions of the semimajor axes of 17 and 100 mas, and the periods of 42 and 100 yr. The white dashed line indicates equal-mass components, and the region above this line is not physical. Blue lines mark 0.08 $M_\odot$, and yellow lines mark the change of grid sizes from 0.01 to 0.002 $M_\odot$. Red boxes highlight the possible systems which can produce the observed perturbations. There is no red box in the 100 mas/42 yrs case because there are no mass combinations consistent with the astrometric data.

Three sets of imaging data provide additional information about the nature of the companion causing the perturbation on LHS 1678. All mass-component mass estimates in this subsection use the Benedict et al. (2016) MLR.

1. The CTIO/SMARTS 0.9 m telescope images used for astrometry also eliminate companions at separations down to 1″ (corresponding to 19.9 au) that are 3.0 $V$ band magnitudes fainter than LHS 1678. This eliminates main sequence companions down to $V \sim 15.5$ and $\mathcal{M}_V = 14.1$, corresponding to a mass of $\sim 0.13\, M_\odot$.
2. The multiepoch speckle imaging from the SOAR HRCam revealed no companions within the limits of those observations (Table 3; Vrijmoet et al. 2022). In particular, no companion was detected at a separation of 0″.15 (corresponding to 2.9 au) down to 2.8 magnitudes fainter than LHS 1678, or to $I \sim 13.1$ and $\mathcal{M}_I \sim 11.7$. The limit at 1″.0 (19.9 au) is 4.9 magnitudes fainter, or to $I \sim 15.2$ and $\mathcal{M}_I \sim 13.8$. There is no reliable mass–luminosity relation available at $I$, but comparisons of $\mathcal{M}_V$ and $\mathcal{M}_I$ for nearby red dwarfs indicate that $\mathcal{M}_I = 11.7$ corresponds to stars with $\mathcal{M}_V \sim 15.3$ and yields a mass limit of $0.12\, M_\odot$ for a companion at a projected separation of $\sim 3$ au. Similarly, $\mathcal{M}_I = 13.8$ corresponds to stars with $\mathcal{M}_V \sim 18.6$, which implies a mass limit of $0.07\, M_\odot$, effectively the end of the stellar main sequence for a companion at 19.9 au.
3. NaCo VLT adaptive-optics imaging did not detect any companions within the limits of those observations. Adopting $Br\gamma$ as equivalent to $K$ band, at separations $\approx 0″.2$ the NaCo image is sensitive to companions with $\Delta K \approx 4$ mag. This corresponds to main-sequence-bound companions at the H-burning mass limit at $\approx 3$ au projected separations. At wider separations of $\gtrsim 0″.5$ the NaCo data are sensitive to companions $\Delta K \lesssim 6.5$ mag. This magnitude difference pushes into the brown dwarf regime where age and luminosity are dependent. An accurate age estimate for LHS 1678 is not available, but the available data indicate the star is not pre-main sequence and may be part of an older galactic population (see Section 3). Adopting an age of 5 Gyr, conservative mass estimates can be made for the limits of bound companions. We adopt this age and use the brown dwarf cooling models of Baraffe et al. (2003) to estimate that the NaCo observations rule out bound companions $\gtrsim 70\, M_{\rm Jup}$ at $\gtrsim 5$ au projected separations. These deep-imaging limits place strong constraints on the allowed parameter space of the astrometric companion. They indicate it is at or below the H-burning mass limit, consistent with the astrometric constraints, and likely closer than $\approx 5$ au in projected separation.

### C.3. Limits Placed by Radial Velocity

The lack of a detection in CHIRON and HARPS radial velocity data (Tables 8 and 6, and Sections 4.3, 5) over the observation baseline is consistent with a low-mass, long-period companion. The 10 day CHIRON campaign and 124 day HARPS campaign are insufficiently long in duration to identify a radial velocity shift caused by a companion with a decades-long orbit. An object, e.g., of mass $0.08\, M_\odot$, with a circular, edge-on orbit around LHS 1678 and a 42 yr period would have a semiamplitude of 1.2 km s$^{-1}$. The semiamplitude would be 0.9 km s$^{-1}$ in the case of a 100 yr period. Both seem easily detectable by CHIRON or HARPS at first glance, but the change in RV would be very gradual. We estimate that such a companion would present itself as a roughly 25 times smaller, $\sim 40$ m s$^{-1}$ shift in RV across the HARPS observation period. We do not detect such a shift. This reaffirms our conclusion that the companion is likely substellar in nature.

### C.4. Could the Companion be a White Dwarf?

In the above discussion, we have assumed that the companion is a low-mass red or brown dwarf, but we must also consider the possibility that the companion is a white dwarf.

To understand which white dwarfs can be eliminated by the imaging observations, we examine those typically found in the solar neighborhood, as outlined clearly in Figure 3 in Subasavage et al. (2017). Most white dwarfs have $\mathcal{M}_V = 10.5$–16.0, although there are rare white dwarfs as faint as $\mathcal{M}_V \sim 16.5$. From the CTIO/SMARTS 0.9 m data, white dwarfs down to $\mathcal{M}_V \sim 14.1$ can be eliminated beyond 1″0, which corresponds to, roughly, the more massive half of the nearby white dwarf population. The SOAR data eliminate companions down to 0″15 brighter than $\mathcal{M}_I \sim 11.7$, but because white dwarfs have $V - I = -0.3$ to $+1.9$, the SOAR data only eliminate the bluest, most-massive white dwarfs. Because the NaCo observations cover the near-IR and white dwarfs are relatively blue, they are not sensitive to (m)any white dwarfs.

However, the radial velocity and astrometry can eliminate the possibility that the companion is a white dwarf. The lowest-mass white dwarfs have masses of $\sim 0.2\, M_\odot$ (Kilic et al. 2007), and at the same distance such objects are much fainter than the value of $\mathcal{M}_V = 11.07$ for LHS 1678. Thus, any such companion would contribute negligible light to the system. As described previously, the HARPS radial velocity nondetection excludes the possibility of a companion more massive than $0.08\, M_\odot$ in 40 to 100 yr edge-on, circular orbits. The astrometry also limits the mass of an unseen companion that contributes negligible light to the system, regardless of whether that companion is a red, brown, or white dwarf. This invisible nature means a white dwarf would contribute as much light to the astrometry as a brown dwarf, and the analysis in Section C.1 that eliminates most companions above $0.08\, M_\odot$ would apply to this one, as well. Thus, we conclude that the companion is very unlikely to be a white dwarf (or a more massive, but even fainter, object such as a neutron star).

### ORCID iDs

Michele L. Silverstein ● https://orcid.org/0000-0003-2565-7909
Joshua E. Schlieder ● https://orcid.org/0000-0001-5347-7062
Thomas Barclay ● https://orcid.org/0000-0001-7139-2724
Benjamin J. Hord ● https://orcid.org/0000-0001-5084-4269
Wei-Chun Jao ● https://orcid.org/0000-0003-0193-2187
Eliot Halley Vrijmoet ● https://orcid.org/0000-0002-1864-6120
Todd J. Henry ● https://orcid.org/0000-0002-9061-2865
Ryan Cloutier ● https://orcid.org/0000-0001-5383-9393
Veselin B. Kostov ● https://orcid.org/0000-0001-9786-1031
Ethan Kruse ● https://orcid.org/0000-0002-0493-1342
Jennifer G. Winters ● https://orcid.org/0000-0001-6031-9513
Stephen R. Kane ● https://orcid.org/0000-0002-7084-0529
Keivan G. Stassun ● https://orcid.org/0000-0002-3481-9052
Chelsea Huang ● https://orcid.org/0000-0003-0918-7484
Andrew Vanderburg ● https://orcid.org/0000-0001-7246-5438
C. E. Brasseur ● https://orcid.org/0000-0002-9314-960X
David Charbonneau ● https://orcid.org/0000-0002-9003-484X
David R. Ciardi ● https://orcid.org/0000-0002-5741-3047
Karen A. Collins ● https://orcid.org/0000-0001-6588-9574
Kevin I. Collins ● https://orcid.org/0000-0003-2781-3207
Dennis M. Conti ● https://orcid.org/0000-0003-2239-0567
Ian J. M. Crossfield ● https://orcid.org/0000-0002-1835-1891
Tansu Daylan ● https://orcid.org/0000-0002-6939-9211
John P. Doty ● https://orcid.org/0000-0003-2996-8421
Courtney D. Dressing ● https://orcid.org/0000-0001-8189-0233
Emily A. Gilbert ● https://orcid.org/0000-0002-0388-8004
Keith Horne ● https://orcid.org/0000-0003-1728-0304
Jon M. Jenkins ● https://orcid.org/0000-0002-4715-9460
David W. Latham ● https://orcid.org/0000-0001-9911-7388
Andrew W. Mann ● https://orcid.org/0000-0003-3654-1602
Elisabeth Matthews ● https://orcid.org/0000-0003-0593-1560
Leonardo A. Paredes ● https://orcid.org/0000-0003-1324-0495
Samuel N. Quinn ● https://orcid.org/0000-0002-8964-8377
George R. Ricker ● https://orcid.org/0000-0003-2058-6662
Richard P. Schwarz ● https://orcid.org/0000-0001-8227-1020
Ramotholo Sefako ● https://orcid.org/0000-0003-3904-6754
Avi Shporer ● https://orcid.org/0000-0002-1836-3120
Christopher Stockdale ● https://orcid.org/0000-0003-2163-1437
Thiam-Guan Tan ● https://orcid.org/0000-0001-5603-6895
Guillermo Torres ● https://orcid.org/0000-0002-5286-0251
Joseph D. Twicken ● https://orcid.org/0000-0002-6778-7552
Roland Vanderspek ● https://orcid.org/0000-0001-6763-6562
Gavin Wang ● https://orcid.org/0000-0003-3092-4418
Joshua N. Winn ● https://orcid.org/0000-0002-4265-047X